\def\draft{0}

\documentclass[11pt]{extarticle}
\usepackage[left=1in,top=1in,right=1in,bottom=1.0in]{geometry} 
\usepackage{cancel}
\usepackage{algorithm2e}
\usepackage[reqno]{amsmath}
\usepackage[T1]{fontenc}
\usepackage{amssymb,amsmath,graphicx,verbatim,url,color,rotating,multirow}
\usepackage{ifthen}

\newtheorem{theorem}{Theorem}[section]
\newtheorem{lemma}[theorem]{Lemma}
\newtheorem{definition}[theorem]{Definition}

\newcommand{\qed}{\nobreak \ifvmode \relax \else
      \ifdim\lastskip<1.5em \hskip-\lastskip
      \hskip1.5em plus0em minus0.5em \fi \nobreak
      \vrule height0.75em width0.5em depth0.25em\fi}

\usepackage{enumitem}

\usepackage{tikz}
\usetikzlibrary{shapes,arrows,calc,positioning,shapes.geometric}
\tikzset{>=latex}

\usepackage{color}

\definecolor{spot}{rgb}{0.6,0,0}
\definecolor{corn}{RGB}{255, 248, 220}
\usepackage[pdftex, bookmarksopen=true, bookmarksnumbered=true,
  pdfstartview=FitH, breaklinks=true, urlbordercolor={0 1 0},
  citebordercolor={0 0 1}, colorlinks=true,
            citecolor=spot,
            linkcolor=spot,
            urlcolor=spot,
            pdfauthor={Privacy Tools Group}
pdftitle={Paper on Privacy Tool}]{hyperref}

\ifnum\draft=1
\newcommand{\AuthorNote}[3]{{\color{#3} {\bf [[#1:}~#2{\bf]]}}}
\else
\newcommand{\AuthorNote}[3]{}
\fi

\usepackage[lofdepth,lotdepth]{subfig}
\usepackage{fullpage}

\newcommand{\thesystem}{\textrm{PSI}\xspace}
\newcommand{\thesystemfull}{a Private data Sharing Interface}
\newcommand{\dataverse}{\textsf{Dataverse}\xspace}
\newcommand{\zelig}{\textsf{Zelig}\xspace}
\newcommand{\tworavens}{\textsf{TwoRavens}\xspace}

\begin{document}
\title{\thesystem ($\Psi$): \thesystemfull\footnote{This work is part of the ``Privacy Tools for Sharing Research Data'' project at Harvard, supported by NSF grant CNS-1237235 and grants from the Sloan Foundation.  This is a working paper describing a vision for work that is still in progress, and is therefore authored by the leadership of the efforts.  Future and accompanying publications that emphasize specific technical contributions will be authored by team members responsible for those contributions.}\\ \normalsize \textsc{(working paper)}}

\author{
Marco Gaboardi\thanks{Department of Computer Science and Engineering, University at Buffalo, SUNY. Work
done in part while at the University of Dundee, UK and visiting the Center for Research on Computation \& Society, John A. Paulson School of Engineering \& Applied Sciences, Harvard University. \texttt{gaboardi@buffalo.edu}.}
\and James Honaker \thanks{\texttt{james@hona.kr}; \url{http://hona.kr}}
\and Gary King \thanks{Albert J.\ Weatherhead III University Professor, Harvard University, Institute for Quantitative Social Science. \texttt{king@harvard.edu}; \url{http://GaryKing.org}}
\and Jack Murtagh\thanks{Center for Research on Computation \& Society, John A. Paulson School of Engineering \& Applied Sciences, Harvard University. \texttt{jmurtagh@g.harvard.edu}.}
\and Kobbi Nissim\thanks{Department of Computer Science, Georgetown University, {\em and} Center for Research on Computation \& Society, John A. Paulson School of Engineering \& Applied Sciences, Harvard University. \texttt{kobbi.nissim@georgetown.edu}.} 
\and Jonathan Ullman\thanks{College of Computer and Information Sciences, Northeastern University.  Work done in part while affiliated with the Center for Research on Computation \& Society, John A. Paulson School of Engineering \& Applied Sciences, Harvard University.  \texttt{jullman@ccs.neu.edu}}
\and Salil Vadhan\thanks{Center for Research on Computation \& Society, John A. Paulson School of Engineering \& Applied Sciences, Harvard University.  Work done in part while visiting the Shing-Tung Yau Center and the Department of Applied Mathematics at National Chiao-Tung University in Taiwan.  Also supported by a Simons Investigator Award.  \texttt{salil@seas.harvard.edu}.}
\and \\
\small with contributions from\\ \small
Nabib Ahmed,
Andreea Antuca,
Brendan Avent,
Jordan Awan,
Christian Baehr,
Connor Bain,
Victor Balcer,\\ \small
Thomas Brawner,
Jessica Bu,
Mark Bun,
Stephen Chong,
Fanny Chow,
Katie Clayton, 
Holly Cunningham, \\ \small
Vito D'Orazio,
Gian Pietro Farina,
Anna Gavrilman,
Benjamin Glass,
Caper Gooden,
Paul Handorff,\\ \small
Raquel Hill,
Alyssa Hu, 
Jason Huang,
Justin Kaashoek,
Allyson Kaminsky,
Chan Kang,
Murat Kuntarcioglu, \\ \small
Vishesh Karwa,
George Kellaris,
Michael Lackner,
Jack Landry,
Hyun Woo Lim,
Giovanni Malloy,\\ \small
Michael Lopiccolo,
Nathan Manohar,
Ross Mawhorter,
Dan Muise,
Marcelo Novaes,
Ana Luisa Oaxaca, \\ \small
Raman Prasad, 
Sofya Raskhodnikova,
Grace Rehaut,
Ryan Rogers,
Or Sheffet,
Adam D. Smith,\\ \small
Thomas Steinke,
Kathryn Taylor,
Julia Vasile,
Clara Wang,
Haoqing Wang,
Remy Wang,\\ \small
Lancelot Wathieu,
David Xiao,
Anton Xue,
and Joy Zheng}

\begin{titlepage}
\maketitle


\begin{abstract}
We provide an overview of the design of \thesystem (``\thesystemfull''), a system we are developing
to enable researchers in the social sciences and other fields to share and explore privacy-sensitive datasets with the strong privacy protections of differential privacy.
\end{abstract}
\end{titlepage}

\section{The Problem}
\label{sec:problem}
Researchers in all empirical fields are increasingly expected to widely share the data behind their published research, to enable other researchers to verify, replicate, and extend their work. Indeed, data-sharing is now often mandated by funding agencies \cite{OMB2013, NIH2003, NSF2014} and journals \cite{AK16, GK13, VH15}. To meet this need, a variety of open data repositories have been developed to make data-sharing easier and more permanent.  The index from the \emph{Registry of Research Data Repositories} surpassed 1500 different repositories in April 2016 \cite{re3data16}.

The largest general-purpose repositories include those that use the open-source \dataverse\ platform \cite{Crosas11, King07} (including Harvard's Dataverse repository, which has, under some measures, the largest repository of social science datasets in the world), CERN's Zenodo, and the commercial Figshare \cite{Singh11}, and Dryad \cite{White08} repositories.

However, many of the datasets in the social and health sciences contain sensitive personal information about human subjects, and it is increasingly recognized that traditional approaches such as stripping ``personally identifying information'' are ineffective at protecting privacy, especially if done by a lay researcher with no expertise in deidentification.  This leads to two problems, one for privacy and one for utility:
\begin{enumerate}[leftmargin=*,itemindent=1.2em,itemsep=.2em]
\item There are numerous data sets, such as surveys, that have been ``deidentified'' via traditional means and increasingly are being 
deposited in publicly accessible data repositories.  As the literature has repeatedly shown, it is likely that many subjects in these surveys can be reidentified by attackers with a moderate amount of background information, and thus their privacy may not be sufficiently well-protected.

\item There are numerous other data sets that are not available at all, or only with highly restrictive and time-consuming provisions.  Such provisions can include a review by the original data depositor---who may no longer be accessible---and/or an Institutional Review Board (IRB), and a lengthy negotiation between institutions on the terms of use.
\end{enumerate}
Thus, an important problem is to develop and deploy methods that can be used to offer greater privacy protections for datasets of the first type, ideally at little or no cost in utility\footnote{Even traditional de-identification techniques have been found to have a significant negative impact on utility~\cite{DariesEtAl14}.}, and enable the safe sharing of datasets of the second type.

Differential privacy~\cite{DMNS06} offers an attractive approach to addressing this problem. Indeed, it provides a formal mathematical framework for measuring and enforcing the privacy guarantees provided by statistical computations. 

The level of privacy protection that differential privacy can offer is described in terms of two privacy loss parameters $\epsilon$ and $\delta$; the smaller they are, the greater the level of privacy. To achieve this greater level of privacy protection, a differentially private algorithm will generally inject a greater amount of random ``noise'' into a statistical computation, thereby yielding less ``accurate'' results.

Using differential privacy enables us to provide wide access to statistical information about a dataset without
worries of individual-level information being leaked inadvertently or due to an adversarial attack.  

There is now both a rich theoretical literature on differential privacy and numerous efforts to bring differential privacy closer to practice, including large-scale deployments by Google \cite{erlingsson2014rappor}, Apple \cite{greenberg2016apple}, and the U.S. Census Bureau~\cite{OnTheMap} (See Section~\ref{sec:previous} for more on previous work.)  However, none of the past work
simultaneously meets all of our desiderata for such a system:
\begin{itemize}[leftmargin=*,itemindent=1em,itemsep=.2em]
\item Accessibility by non-experts:~researchers in the social sciences should be able to use the system to share and explore data with no involvement from experts in data privacy, computer science, or statistics. 
\item Generality: the system should be applicable and effective on a wide variety of heterogeneous datasets, as opposed to being tailored for a particular data source or domain.
\item Workflow-compatibility: the system should fit naturally in the workflow of its users (e.g. researchers in the social sciences), and be positioned to offer clear benefits (e.g. more access to sensitive data or less risk of an embarrassing privacy violation) rather than being an impediment.
\end{itemize}

\section{Our Contribution: \thesystem}
In this paper, we provide an overview of \thesystem (``\thesystemfull''), a system we have developed to enable researchers in the social sciences and other fields to share and explore privacy-sensitive datasets with the strong privacy protections of differential privacy.  It is designed to achieve all of the desiderata mentioned in Section~\ref{sec:problem} (Accessibility for Non-Experts, Generality, and Workflow-compatibility).  Unique features of \thesystem\ include:

\begin{itemize}[leftmargin=*,itemindent=1em,itemsep=.2em]
\item None of its users, including the {\em data depositors} who have privacy-sensitive data sets they wish to share and the {\em data analysts} who seek to analyze those datasets, are expected to have  expertise in privacy, computer science, or statistics.  Nevertheless, \thesystem\ enables them to make informed decisions about the appropriate use of differential privacy, the setting of privacy loss parameters, the partitioning of a privacy budget across different statistics, and the interpretation of errors introduced for privacy.
\item It is designed to be integrated with existing and widely used data repository infrastructures, such as the \dataverse\ project \cite{Crosas11, King07}, as part of a broader collection of mechanisms for the handling of privacy-sensitive data, including an approval process for accessing raw data (e.g. through IRB review), access control, and secure storage.   Consequently, \thesystem\ can initially be used to {\em increase} the accessibility of privacy-sensitive data, augmenting rather than replacing current means for accessing such data, thereby lowering the adoption barrier for differential privacy. 
\item Its initial set of differentially private algorithms were chosen to include statistics that have wide use in the social sciences, and are integrated with existing 
statistical software designed for lay social science researchers, namely the \zelig~\cite{Choirat15} package in R and the \tworavens~\cite{honaker2014statistical} graphical data exploration interface. 

We have developed a prototype of the system. Integration with \dataverse is ongoing and will be live in the near future. The features of \thesystem\ described in this paper are at differing stages of completion. Some have been implemented and thoroughly evaluated in both user tests and replication experiments (see Sections \ref{sec:uiux} and \ref{sec:evaluation} , respectively), some have been implemented but have not undergone user testing either intentionally to reduce participant fatigue or because they were still in progress at the time of testing, and others are actively in development. Throughout the text we indicate which aspects of the system are implemented and which are in development, and the sections on user testing and replication experiments describe which pieces of the tool underwent those evaluations. 
\end{itemize}

A preliminary prototype of \thesystem\ is available at \url{http://privacytools.seas.harvard.edu/psi}.  It does not yet incorporate all of the planned features described in this paper, as a number of them are still under development.  The purpose of this paper is to describe the {\em design} of \thesystem, and initiate a discussion about the choices made and possible alternatives.

\section{A Motivating Story}
\label{sec:story}
Consider a social scientist, Alice, who is studying the relationship of health status to political participation.  Since the rules surrounding health insurance have been a recent focus of political debate, the researcher is interested to see if individuals with health problems have become more engaged in the political process.  She searches the catalog of the tens of thousands of datasets archived in a Dataverse data repository and locates several that may contain information to test her hypothesis.  Some are broad surveys of attitudes and behavior that contain hundreds of questions across many domains,\footnote{For example the General Social Survey, National Election Study, or Cooperative Congressional Election Study.} so likely contain only a couple of questions on voting turnout or a couple of questions on health status.  Some are more focused studies that survey patient populations, with richer questions about their medical issues and questions to judge the impacts on their lives and opinions, or longitudinal studies that revisit these participants repeatedly over a long time scale.  She might also locate detailed time diary studies, where respondents agree to provide extensive recordings of how they spend their time each day, spaced with periodic surveys, that might even include biological surveys of cortisol and other hormone levels from saliva.  

The broad surveys are generally available for public download from the repository; however, even here, geographic variables such the state of residence, are only available in a special version that is closed to the public.\footnote{For example, to get the state of residence of respondents, the General Social Survey requires a signed contract with the researcher, a thorough description of the research to be conducted on the data, a fee of \$750, approval from the researcher's IRB, and construction of a data protection plan that generally requires non-networked computers set up in a room with secured limited access \cite{NORC}.}  The focused studies may or may not be publicly available for download depending on the original data depositor's wishes, agreements with the depositor's Institutional Review Board (IRB), and whether the data touches on vulnerable protected populations (such as children, felons, minorities, the disabled) or data for which there are federal regulations (such as health care and student records).  The time diary studies, because of their rich description of individuals and potential for reidentification, will almost certainly be stored but unavailable to the researcher.

In summary, as the datasets become richer, and more likely to directly test her hypotheses, they are more likely to be unavailable for download.  So our researcher can use the data easily available and test the broadest implications of her theory with somewhat crude proxy variables for the items of interest.  Or she can apply to gain access to sensitive data that can allow her to directly test all of her hypotheses in nuance.  Each application delays the research project, is costly in terms of researcher time and IRB resources, and commonly requires participation or approval by the original data depositor and relevant IRBs, all of which in turn may need to be facilitated by communication through the repository staff curators. Even if a researcher is committed to applying for access to private data, the available descriptions of the data are often insufficient to judge which of the potential closed datasets contain the best or most relevant data for the researcher's purposes.  Many of these IRB applications may turn out to be costly lost efforts once access is finally granted.  For example, there may be an insufficient number of data points of the type required, or the variables are not measured in the manner expected, or the time period is wrong.

Imagine instead that the repository enabled Alice to run exploratory statistical procedures on the datasets in Dataverse, so as to learn which datasets would be useful to her research, but only returning statistical answers that are differentially private, so as to uphold the repository's ethical and legal responsibilities to safeguard sensitive data.  She could investigate all the sensitive datasets that are closed for download, and use the noisy statistical answers to learn which datasets have information useful for her research.
Our \thesystem\ system is built to allow such immediate exploratory access to these closed files and to reduce the wasted effort of researchers, data depositors, repositories and IRBs coming from applications to access to datasets that eventually prove to not be useful to the applicant.

\begin{figure}
  \centering
  \fbox{\includegraphics[scale=.20]{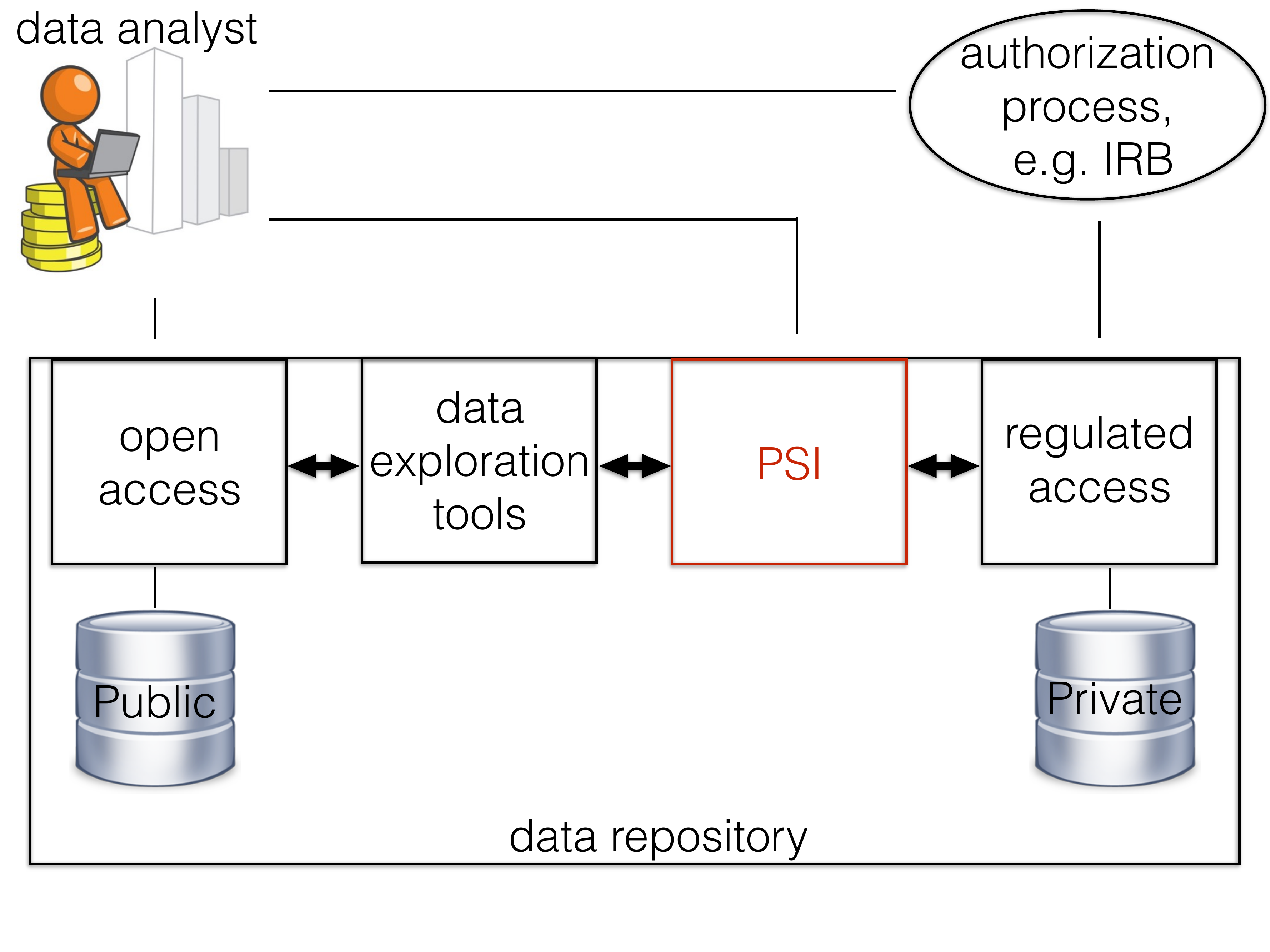}}
    \caption{Overview of a data analysis repository using \thesystem.}
\label{fig:overview}
\end{figure}

The situation we have described above is schematically represented in Figure~\ref{fig:overview}. When a data analyst has access to a data repository infrastructure like Dataverse, they currently have only two options: either using the publicly available data which are offered with open access, or going through an authorization process that may be lengthy and costly.  With \thesystem\ there will be a valuable
alternative: accessing the sensitive data for data exploration. To enhance this opportunity, \thesystem\ is designed to be naturally integrated  with both the data repository infrastructure and data explorations tools. 

\section{Differential Privacy}
Differential privacy is a formal mathematical framework for measuring the privacy guarantees provided by statistical computations.   Consider an algorithm $M$ that takes a dataset $x$ as input and performs a randomized computation to produce an output $y$. Informally, differential privacy requires that if we change any one individual's data in $x$, then the distribution of $y$ does not change much.  Intuitively, this means that
each individual's data is hidden from an adversary that views the output $y$.

To make this intuition precise, we need to define what we mean by ``one individual's data,'' and provide a measure of how much the distribution of $y$ is allowed to change.   For the former, a typical choice is to consider
datasets $x$ that consist of $n$ records, where we think of each record as consisting of one individual's data, and the sample size $n$ is public (not sensitive information).  
We call two datasets $x$ and $x'$ {\em neighbors} if they agree in all but one record (i.e. $x'$ is obtained from $x$ by changing one individual's data).  Then the formal definition of differential privacy is as follows:

\begin{definition}[Differential Privacy, \cite{DMNS06,DworkKeMcMiNa06}]
For parameters $\epsilon\geq 0$ and $\delta\in [0,1]$, a 
randomized algorithm $M$ is {\em $(\epsilon,\delta)$-differentially private} if 
for every two neighboring datasets $x,x'$ and every
set $S$ of outputs, 
$$\Pr[M(x)\in S] \leq e^{\epsilon}\cdot \Pr[M(x')\in S]+\delta,$$
where the probabilities are taken over the randomization of the algorithm $M$.
\end{definition}

The level of privacy protection is governed by the two {\em privacy loss parameters} $\epsilon$ and $\delta$; the smaller they are, the closer the distributions of $M(x)$ and $M(x')$ are, and hence the greater the level of privacy.    Typically, $\epsilon$ is taken to be a small constant such as $.1$, whereas $\delta$ is taken to be very small, like $2^{-30}$.

The way that differentially private algorithms for statistical analysis are often designed are by carefully introducing a small amount of random noise into non-private algorithms for the same analyses. The more noise that is introduced, the greater the level of privacy protection (i.e.~a smaller $\epsilon$ and/or $\delta$). However, less noise produces a more accurate and useful analysis.  Thus differentially private algorithms offer a privacy-utility tradeoff.

In order to limit the amount of noise required, most differentially private algorithms for statistical analysis require the number of data points and the explicit range of the different data variables. A simple example is the mean of a numeric variable, the amount of noise needed to guarantee differential privacy for it is proportional to its range divided by the number of data points.

By now, there is a large literature giving differentially private algorithms for a wide variety of data analysis tasks.   Often, these algorithms are accompanied by a theoretical analysis showing that their performance converges to that of the non-private algorithm as the sample size $n$ tends to infinity.   However, such asymptotic performance guarantees
do not necessarily translate to good performance at a specific finite sample size, and thus a great deal of work remains to be done to engineer differentially private algorithms to be useful in practice.

In addition, one typically does not want to run just one analysis on a dataset, but rather a large collection of analyses.   Fortunately, differentially privacy satisfies a variety of {\em composition theorems} showing that the privacy protection degrades gracefully when we run many differentially private algorithms.   For example:

\begin{theorem}[Basic Composition~\cite{DMNS06,DworkKeMcMiNa06}]
\label{thm:basiccomp}
Let $M_1,\ldots,M_k$ be randomized algorithms where $M_i$ is $(\epsilon_i,\delta_i)$ differentially private for $i=1,\ldots,k$.   Then the algorithm $M(x)=(M_1(x),\ldots,M_k(x))$ that runs each of the $M_i$'s using independent coin tosses is
$(\sum_i \epsilon_i,\sum_i \delta_i)$ differentially private.
\end{theorem}

If we want to achieve a global, overall level of privacy protection given by $(\epsilon_g,\delta_g)$, we can think of the pair as a ``privacy budget'' to be spent on different analyses $M_i$ we want to run. We can spend more of this budget on a specific analysis $M_i$ (i.e.~take $\epsilon_i,\delta_i$ smaller), but this will consume more of our budget, leaving less for the other analysis if we want to ensure that $\sum_i \epsilon_i\leq \epsilon_g$ and $\sum_i \delta_i\leq \delta_g$.

There are better bounds on the composition of differentially private algorithms than the
simple summing bound given above~\cite{DRV10,KOV15,MurtaghV16}, but they still have the same budget-like effect---a larger $(\epsilon_i,\delta_i)$ (i.e.~higher accuracy, lower privacy) for one computation requires reducing the $\epsilon$ and $\delta$ values for other computations in order to maintain the same overall level of privacy.

\section{Previous work} \label{sec:previous}
Most of the previous work to bring differential privacy to practice can be partitioned into the following categories:
\begin{itemize}[leftmargin=*,itemindent=1em]
\item \textit{Programming languages and systems:}   here the goal is to make it easier for users to write programs that are guaranteed to be differentially private, either by composition of differentially private building blocks~\cite{McSherry09,ReedP10,Haeberlen11}, using general frameworks such as ``partition-and-aggregate'' or ``subsample-and-aggregate''~\cite{NRS07} to convert non-private programs into differentially private ones~\cite{RSKSW10,MTSSC12}, or by formal verification from scratch~\cite{conf/popl/BartheKOB12}.  On one hand, these methods provide much more generality than we seek---our target users are not programmers, and it will already be very useful to provide them with a small, fixed collection of differentially private versions of statistical computations that are common in the social sciences.  On the other hand, most of these tools do not provide much guidance for a lay user in deciding how to partition a limited privacy budget among many statistics or analyses he or she may want to run, or how to interpret the noisy results given by a differentially private algorithm. 

In contrast to the other tools mentioned above, GUPT~\cite{MTSSC12} does enable a user to specify fine-grained accuracy goals and automatically converts these into privacy budget allocations, in a similar spirit to our privacy budgeting tool (described later).   However, GUPT is limited to differentially private programs obtained via the subsample-and-aggregate framework, whereas our tool has no such restriction, and can be extended to include arbitrary differentially private algorithms.
Moreover, our tool allows the privacy budget allocation to be interactively adjusted by users, and supports optimal composition theorems for differential privacy~\cite{MurtaghV16}.

\item \textit{Optimization for specific data releases:} there have been several successful applications of differential privacy to very specific and structured sources of data like commuter patterns~\cite{OnTheMap}, mobility data~\cite{DP-where}, client-side software data~\cite{EPK14},  and genome-wide association studies~\cite{JiangEtAl14}

Here differential privacy experts carefully optimize the choice of differentially private algorithms and the partitioning of the privacy budget to maximize utility for the particular data source.   In the context of a broad data repository in the social or health sciences, the collection of data sources and the structure of the datasets is too heterogenous to allow for such optimization.  And it is not scalable to have a differential privacy expert manually involved in each instance of data sharing.

\item \textit{Optimization and evaluation of specific  algorithms:}  there is a vast literature
on the design of differentially private algorithms for specific data analysis tasks, including substantial experimental work on comparing and optimizing such algorithms across a wide range of datasets. As an example, the recent work on DPBench~\cite{HayMMCZ15} provides a thorough comparison of different algorithms and different ways of optimizing them.

Such work is complementary to ours.  Algorithms that perform well in such evaluation are natural candidates to add to our library of differentially private routines, but such evaluation does not address how to budget the privacy allocated to this one algorithm against many other analyses one might want to run on the same dataset or more generally how to enable lay users to make appropriate use of differential privacy. Moreover, our use case of a general-purpose social science data repository guides the choices of which algorithms to implement, the measures of accuracy, and the methods for evaluation, as discussed in the later sections.
\end{itemize}

There are also a number of deployed systems that provide query access to sensitive data, using heuristic approaches to protect privacy.   These include systems for querying clinical health data~\cite{STRIDE09,SHRINE13},
education data~\cite{NAEP11}, genomic data~\cite{Tryka01012014}, and Census data~\cite{FactFinder}. However, the lack of rigorous privacy guarantees raises a genuine risk, as illustrated by attacks on the Israeli
Census query system~\cite{Ziv13}, on genomic data~\cite{Homer+08,ErlichN14} 
and more generally on
releases of aggregate statistics~\cite{DinurNi03,DworkSmStUl16}.  (Some of the aforementioned systems address this concern by limiting access to a more trusted set of users.)

\section{Incentives for use} \label{sec:incentives}
Differential privacy has sometimes been critiqued 
for its cost in utility (coming from the noise introduced in statistics), thus one might wonder what would motivate researchers to use it in place of the current data-sharing ecosystem.
We see at least three different scenarios in which differential privacy can provide a clear benefit over current approaches.

\begin{itemize}[leftmargin=*,itemindent=1em]
\item (``DP works great'') In some circumstances, the results of differentially private analyses are virtually indistinguishable from non-private analyses.  Currently, this tends to be the case when the number $n$ of samples is large, the data is low-dimensional, and the analyses to be performed are relatively simple and few in number.  In such cases, the greater privacy protections of differential privacy come essentially for free.  As both theoretical and applied work on differential privacy advances and data gets ``bigger'' ($n$ gets larger), we can expect an increasingly large set of data-sharing circumstances to fall in this scenario. 

\item (``Access is wide'')  When we wish to make sensitive data available to an extremely wide community (for example, when allowing public access), we should be increasingly concerned about attacks from individuals with malicious intent.  Such adversaries can include ones who have extensive knowledge about a particular data subject that can be exploited as background information.  Thus, the strong protections of differential privacy, which remain meaningful regardless of an adversary's background information, are attractive.

\item (``Data is currently unavailable'')  For data that is currently unavailable except possibly through restrictive and time-consuming provisions, {\em any} useful statistical information that differential privacy can offer is a benefit to utility, even if it does not fall in the ``DP works great'' category.   In particular, DP can offer the possibility of rough, exploratory analysis to determine whether a dataset is of sufficient interest to go through the process of applying for access to the raw data.
\end{itemize}
The architecture of \thesystem\ is designed to support all three of these scenarios.  
In the near term, we expect the third scenario, namely enabling exploratory analysis of data that is currently unavailable, to be the one where \thesystem\ is most frequently used.  In this scenario, \thesystem\ can provide a clear utility benefit, can be applied with the modest sample sizes that are common in social science, and does not require an extensive library of highly optimized and sophisticated differentially private algorithms. 
However, \thesystem\ is extensible to incorporate such a library in the future, and we hope that eventually it will be used more often in the other two scenarios as well, providing high-utility and privacy-protective access to data that is currently shared in a less safe manner \cite{crosas2015automating}.

In the future, another potential incentive for the use of a differentially private data analysis system like \thesystem\ is the automatic protection that differential privacy provides against false discovery, allowing analysts to perform adaptive data exploration (without ``preregistration'') and still have confidence that the conclusions they draw are statistically valid~\cite{DworkFeHaPiReRo15,BassilyNiSmStStUl16}. 

We note that sometimes researchers do not wish to share their data, and are only using privacy as an excuse.   A system like \thesystem\ can help eliminate the excuse.   Still, other external incentives may be needed (such as from the research community, funding agencies, or journals) to encourage sharing of data.

\medskip

\subsubsection*{On exploratory analysis}  Since it is
our initial goal for the use of \thesystem, we elaborate on what we mean by supporting ``exploratory data analysis.''  This term generally refers to a wide-ranging set of techniques to empirically learn features of data by inspection, and familiarize oneself with the nature of the data, or discover apparent structure in the data \cite{Tukey77}.  It is inspection and discovery not driven by theory or modeling.
In our setting of a social science data repository, 
we envision at least two uses for exploratory
analysis.  For lay-users (e.g. members of the general public), exploratory analysis can be a way to satisfy curiosity and discover interesting facts for situations where a statistically rigorous analysis may not be necessary (e.g. for a high-school project).
For a social science researcher, the goal of exploratory analysis can be 
to determine which of the many datasets in the repository are of most interest, so that the researchers only invest their time and effort in applying for raw access to those datasets.   Any final analyses they wish to perform and publish could then be done on the raw data, not through the differentially private interface.  
This more modest near-term goal for \thesystem\  compensates for the fact that we cannot perform the kinds of optimizations that might be done if we had a differential privacy expert involved in each instance of data sharing.

\section{Actors and Workflow}
\label{sec:actors}

We have three different kinds of actors in \thesystem: data depositors, data curators, and data analysts. Each of them has a different role and different requirements. We represent them,  their interaction  and the threat model we consider, in terms of trust for different actors, in Table~\ref{tab:trust}. We now detail the roles and the expected expertise for each of them.

\begin{description}[leftmargin=*,itemindent=.5em, listparindent=\parindent]
\item[Data depositors.]  These are users that come to deposit their privacy-sensitive dataset in a data repository, and may wish to make differentially private access to their dataset available.  
By interacting with the system, the data depositor supplies basic information about the dataset (e.g. the types and ranges of the variables), sets the overall privacy loss parameters, selects an initial set of differentially private statistics to calculate and release, and determines how the remaining privacy budget will be partitioned among future data
analysts.

Data depositors are the ones with the initial ethical and/or legal responsibility for protecting the privacy of their data subjects, and they (or their institutions) may be liable if they willfully violate their obligations.   Thus, they can be trusted to follow instructions (if not onerous or confusing) and answer questions truthfully to the best of their knowledge.  On the other hand, they cannot be assumed to have expertise in differential privacy, computer science, or statistics, so any questions that involve these areas are explained carefully in the system.   

\item[Data curators.] 
These are the data-repository managers that maintain the hardware and software on which
\thesystem\ runs and the accompanying data repository infrastructure (e.g.~\dataverse) and associated statistical tools (e.g.~\zelig\ and \tworavens).  They are trusted, and
indeed may also have legal obligations to ensure that the repository does not violate the privacy protections it claims to offer through tools such as \thesystem. 
Data curators 
can be assumed to have expertise in IT systems administration and data stewardship \cite{Goodman14} and archiving \cite{Altman09}, and can be trained to have at least a modest background in statistics and differential privacy.   But they are few in number, and cannot be actively involved in most instances of data sharing or data exploration.   Thus \thesystem\ is designed to be sufficiently automated to enable data depositors and data analysts to safely use it on their own.

Data curators would also be responsible for deciding whether to accept new differentially private routines into the library used by \thesystem\ and correcting bugs or security flaws found in existing routines.   These can be difficult tasks even for experts in differential privacy.  Thus, in a future version of the system, it would be of interest to minimize the amount of trusted code, and have tools to formally verify the remaining components (both original components and later contributions), along the lines of
the programming languages tools described in Section~\ref{sec:previous}.

\item[Data analysts.]   These are users 
that come to access sensitive datasets in the repository, often with the goal of data exploration as discussed in 
Section~\ref{sec:incentives}.
They will have access to all of the differentially private statistics selected by the data depositor, as well as the ability to make their own differentially private queries (subject to staying within the overall privacy budget, as discussed more below).  

We envision at least two tiers of trust for data analysts once the system has active users. \thesystem\ can make access available to a very wide community of analysts (e.g. the general public), in which case the analysts are considered completely {\em untrusted}.
Alternatively (or additionally), we can restrict to a
set of analysts that are identifiable (e.g. as registered users of the data repository), with some accountability (e.g. through their verified affiliation with a home institution).   Such analysts may
be considered {\em semi-trusted}, as we can assume that they will follow basic terms of use to not abuse the system in certain ways. Specifically, we will assume that semi-trusted users will not collude to compromise privacy, and will not create phony accounts.  (This will enable us to provide greater utility for such users, as discussed in Section~\ref{sec:budgeting}.)\\
 \begin{table}[t] 
 \centering
 \begin{minipage}[t]{.4\linewidth}
\vspace{0pt}
\centering
\includegraphics[width=\linewidth]{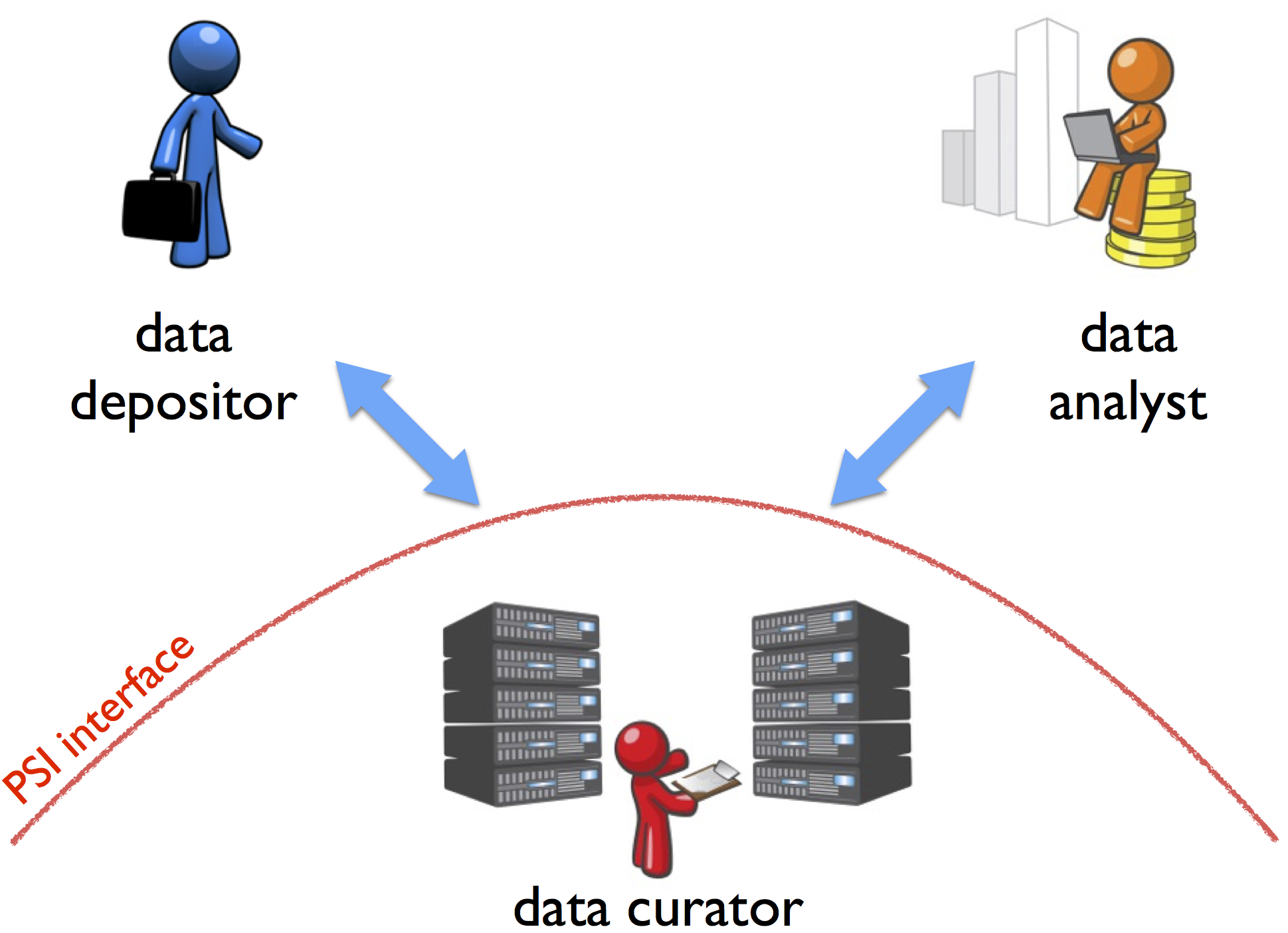}
\end{minipage}%
\ \
\begin{minipage}[t]{.4\linewidth}
\centering
\scalebox{.99}{\begin{tabular}[t]{|c|c|c|}
    \hline
\textbf{Actors} & \textbf{Level of} &\textbf{DP}\\
&\textbf{trust}&\textbf{expertise}\\
\hline\hline
data curators & trusted & modest \\
\hline
data depositors & trusted & none \\
\hline
data analysts & semi-trusted & none\\
 (restricted)&&\\
 \hline
data analysts  & untrusted & none\\
(general public)&&\\
\hline
\end{tabular}}
\end{minipage}
\caption{Actors and their level of trust and required expertise.}
\label{tab:trust}
 \end{table}
\end{description}


\section{Pedagogical Materials}
In order to enable \thesystem\ to be used by empirical researchers without expertise in privacy, computer science, or statistics, we have
prepared pedagogical materials explaining differential privacy in an intuitive but accurate manner, with a minimum of technical terminology and notation. These materials are meant to be sufficient for data depositors and data analysts to understand and make appropriate choices in using \thesystem, such as those described in the forthcoming sections.  
Data depositors require more background material than data analysts, as the former are concerned with the privacy protections afforded to their data subjects, whereas the latter only need to understand the impact of the system on their analyses (namely, that results will be less accurate or statistically significant than would be obtained on the raw data, and that there is a limited ``budget'' of queries that they can perform). 

Relevant extracts of the pedagogical materials are offered in \thesystem\ at each decision point, and can also be included when describing data-sharing plans to Institutional Review Boards (IRBs). In addition, members of our team have started to develop rigorous arguments showing that differential privacy should be deemed to satisfy certain legal obligations of privacy protection, which can also be used to reassure data depositors, data curators, and IRBs that differential privacy is a sufficiently strong form of protection. For example, the combined legal and technical analysis in \cite{bridging} provides an argument that, when applied to educational data, differentially private computations satisfy the requirements of the Family Educational Rights and Privacy Act of 1974~\cite{ferpa}.

As discussed in Section~\ref{sec:actors}, we assume that data curators have expertise in IT systems administration and data stewardship, and at least a modest background in statistics and differential privacy.  Thus, they do not
need any specialized pedagogical materials other than a thorough documentation of the system.

\section{Privacy Budget Management}
\label{sec:budgeting}
\begin{figure*}[htb]
    \includegraphics[width=\textwidth]{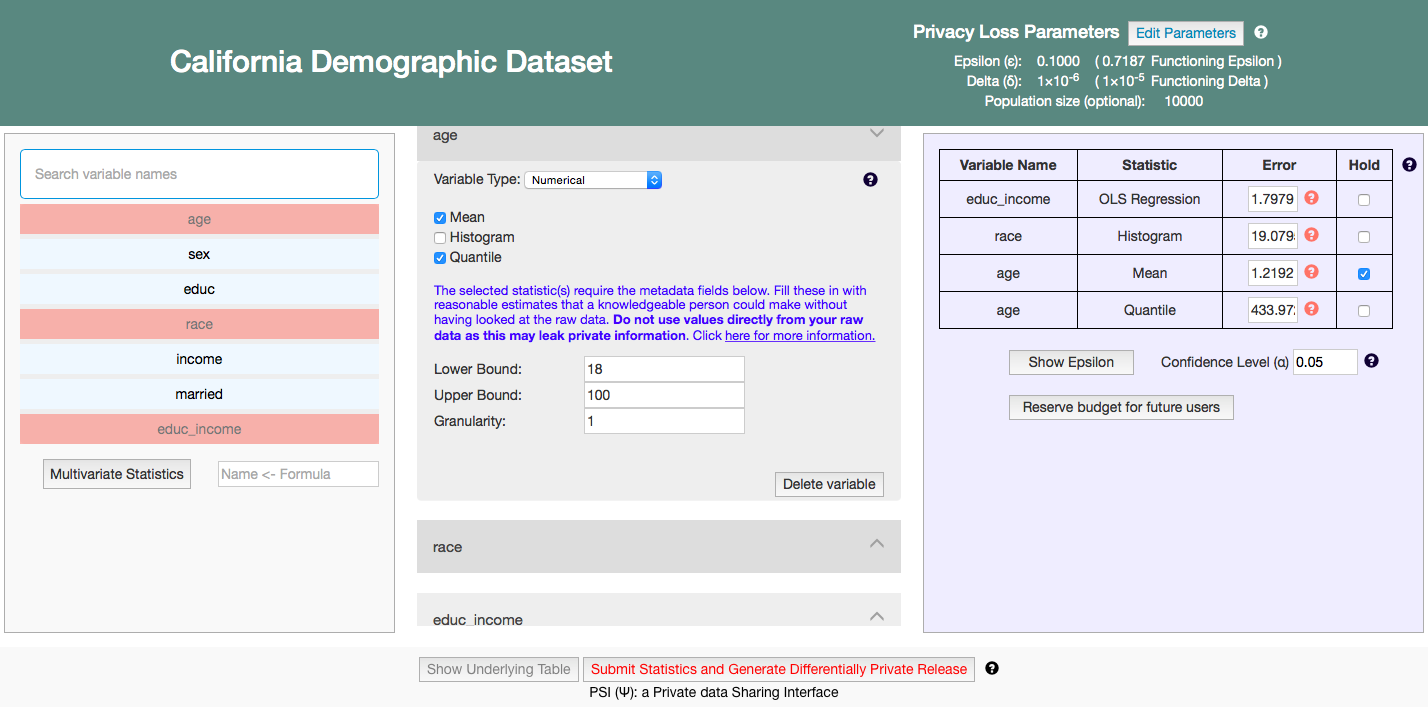}
    \caption{\textbf{\thesystem budgeting interface:} \it The left panel shows variables in the dataset; middle panel elicits necessary metadata from depositors; right panel displays selections with a priori error estimates, tailored to each statistic. For example, the error on the race histogram indicates the released count for each bin will differ from the true count by at most 19.079 people with probability .95.  
The upper right corner displays the privacy loss parameters and the functional boosts from secrecy of the sample.
    } \label{fig:interface}
\end{figure*}

One of the challenges in enabling non-experts to use differential privacy is that it can be difficult to understand the implications of different selections of the privacy loss parameters (namely $\varepsilon$ and $\delta$), both in terms of privacy and utility, especially when these need to be distributed over many different statistics to be computed. To address this issue, \thesystem\ is designed to expose these implications to the user, in easy-to-understand terms, and is accompanied by a variety of simple explanations of differential privacy and its parameters that are shown to the user at relevant times. We have developed a privacy budgeting tool that guides users through judicious choices of global privacy loss parameters, lets users select statistics to release, automatically distributes the privacy budget across the chosen statistics, and exposes the resulting privacy-accuracy tradeoffs (see Figure~\ref{fig:interface}.) 

\begin{description}[leftmargin=*,itemindent=.5em, listparindent=\parindent]
\item[Global privacy loss parameters:] 
The data depositor, who carries the initial responsibility for protecting the privacy of her data subjects, is charged with setting the overall (``global'') privacy loss parameters $\epsilon_g,\delta_g$ for her dataset (seen at the top right of Figure~\ref{fig:interface}).   To enable this choice, we provide intuitive (but accurate!) explanations of the meaning of each of these privacy loss parameters, and give recommended settings based on the level of sensitivity of a dataset (e.g. corresponding to an institution's established research data security levels, such as \cite{HarvardDCT}
or the similar categories in the DataTags system that integrates with \thesystem~\cite{SCB15}). $\delta_g$ is easily explained as the probability of arbitrary leakage of information, like the probability of an adversary breaking an encryption scheme, and thus should be set to be extremely small, like $2^{-30}$.   For the main privacy loss parameter, $\epsilon_g$, we explain it with a table comparing an adversary's posterior belief that a data subject has a sensitive trait to the posterior belief had the subject opted out of the study. \thesystem\ also confirms with the data depositor that each individual subject's data corresponds to one row of the uploaded dataset (so that the per-row protections of differential privacy translate to per-subject protections). 

\item[Secrecy of the sample:]
The data depositor is asked whether the dataset is a  random sample from a larger population, and whether the choice of this sample has been kept confidential. If so, a useful lemma in differential privacy known as ``secrecy of the sample'' allows for an effective savings in the privacy loss parameters corresponding to the ratio of sizes between the dataset and the larger population.\footnote{\url{https://adamdsmith.wordpress.com/2009/09/02/sample-secrecy/}.}  This means that correspondingly greater utility can be provided for the same level of privacy protection.  
(To account for the fact that, in practice, population samples are typically not perfectly
random, the depositor is instructed to conservatively estimate the overall population size.) 

\smallskip

\begin{lemma}[Secrecy of the sample~\cite{KLNRS08, AdamSecrecyPost}]
Let $M$ be an $(\epsilon,\delta)$-differentially private algorithm for datasets of size $n$. Let $M'$ be a randomized algorithm that takes as input a dataset $D$ of size $m\geq n$, and then runs $M$ on a dataset $D'$ obtained by selecting a uniformly random subset of $D$'s records of size $n$. Then, $M'$ is 
$((e^\epsilon-1)\cdot (n/m),\delta\cdot (n/m))$-differentially private.
\end{lemma}

\smallskip

In the application of this lemma in \thesystem, $D'$ represents a dataset that is being deposited in the repository, $D$ represents a larger population from which $D'$ was (randomly) drawn, and $M$ represents the differentially private statistics computed by \thesystem\ on  $D'$. Note that in typical applications of differential privacy, $\epsilon$ is a small constant and therefore $(e^\epsilon-1)\cdot n/m\approx \epsilon\cdot n/m$. In concrete applications, especially in the social sciences, this lemma permits large savings in the privacy budget. For this reason, we integrate this property in the budgeting interface (See Figure~\ref{fig:interface}).

\item[Budgeting among different statistics:]
 Once global privacy loss parameters have been determined, users can select variables from their dataset (from the left-hand panel of the budgeting interface, Figure~\ref{fig:interface}) and choose statistics to release about those variables from \thesystem's library of differentially private algorithms.  At this stage, there is still the challenge of how the global privacy loss parameters should be distributed among the different statistics to be computed.  That is, for each statistic to be computed, we need to select privacy loss parameters (i.e. set $\epsilon_i$ and $\delta_i$ for statistic $i$) and then apply composition theorems to ensure that globally, we achieve $(\epsilon_g,\delta_g)$ differential privacy.   
 
 This leaves the question of how a user should select individual privacy loss parameters $\epsilon_i$ (and $\delta_i$).  The larger the value of $\epsilon_i$ is taken, the more utility we obtain from the $i$'th statistic, but this leaves less of the global privacy budget remaining for the other statistics.  Since some statistics a user is computing may be more important than others, and different differentially private algorithms have different privacy-utility tradeoffs, the ``best'' use of the privacy budget is likely to involve a non-uniform distribution of the $\epsilon_i$'s.   

To enable users to determine this partition without requiring that they be privacy experts, \thesystem\ automatically assigns initial privacy loss parameters to each chosen statistic.  Similarly to
GUPT~\cite{MTSSC12}, \thesystem\ then
exposes the privacy-accuracy tradeoffs to the user (see the summary table in the right-hand panel of Figure~\ref{fig:interface}.) Rather than adjusting the individual privacy loss parameters $\epsilon_i$, the user can instead modify the ``accuracy'' that will be obtained for different selected statistics (presented as, for example, the size of 95\% confidence intervals; see further discussion in the next section).   For each differentially private algorithm in \thesystem, there are accompanying functions that translate between the privacy loss parameters and a measure of accuracy (also depending on other metadata, such as the range of variables involved and the dataset size $n$).  These functions are used by the privacy budgeting tool to translate the accuracy bounds into individual privacy loss parameters and ensure that the global privacy loss parameters are not exceeded.   

\item[Optimal composition:]
To ensure that we get the most utility out of the global privacy budget, we use the Optimal Composition Theorem \cite{MurtaghV16}, which in fact was developed for the purpose of our privacy budget tool. This characterizes the optimal value for the global privacy budget $\epsilon_g$ (for each possible $\delta_g\in [0,1)$) when composing $k$ algorithms that are $(\epsilon_i,\delta_i)$-DP. 

\smallskip 

\begin{theorem}[Optimal Composition Theorem, \cite{MurtaghV16}]
\label{thm:optcomp}
Let $M_1,\ldots,M_k$ be randomized algorithms where $M_i$ is $(\epsilon_i,\delta_i)$ differentially private for $i=1,\ldots,k$ and let $\delta_g\in [0,1)$. Then the algorithm $M(x)=(M_1(x),\ldots,M_k(x))$ that runs each of the $M_i$'s using independent coin tosses is $(\epsilon_g,\delta_g)$ differentially private for the least value of $\epsilon_g$ satisfying the following inequality:

$$\frac{1}{\prod_{i=1}^k (1+e^{\epsilon_i})}\cdot
\sum_{S\subseteq\{1,\ldots,k\}} \max  \left\{ e^{\sum_{i\in S} \epsilon_i}-e^{\epsilon_g} \cdot e^{\sum_{i\not\in S} \epsilon_i},0 \right\}$$
$$\leq 1-\frac{1-\delta_g}{\prod_{i=1}^{k}(1-\delta_i)}$$
\end{theorem}

\smallskip
While the Basic Composition Theorem gives an upper bound on the degradation of privacy under composition, the above theorem is optimal in the sense that for every set of privacy loss parameters, $(\epsilon_i,\delta_i)$ for $i\in\{1,\ldots,k\}$ and $\delta_g$ there exists a set of algorithms $M_1,\ldots, M_k$ that are $(\epsilon_i,\delta_i)$ differentially private, respectively, whose composition achieves $(\epsilon_g, \delta_g)$ differential privacy \emph{exactly}. 

For even moderate values of $k$, the optimal composition theorem can provide substantial savings in the privacy budget over the other composition theorems in differential privacy. In an effort to maximize utility for users, the budgeting interface uses an implementation of Theorem \ref{thm:optcomp} to apportion a global epsilon value across several statistics. Since the Optimal Composition Theorem is infeasible to compute in general, we use an efficient approximation algorithm that still outperforms the alternative composition theorems \cite{MurtaghV16}. 

\item[Budgeting among different actors:]
Recall that the selection of differentially private statistics to be computed is done both by the data depositor, who selects an initial set of statistics that will be all who access the dataset, and by individual data analysts, who may be carrying out novel explorations of their own conception.
The privacy budgeting tool described above is designed to support both types of actors (with slightly different settings for each to reflect their different roles and level of trustworthiness).   The data depositor is tasked with deciding how much of the global privacy budget $\epsilon_g$ to reserve for future data analysts.
For example, if the data depositor uses up $\epsilon_d$ units of privacy for the statistics she chooses to release, then at least $\epsilon_a = \epsilon_g-\epsilon_d$ units of privacy will be left for the future analysts.  ($\epsilon_a$ might actually be larger, since composition theorems for differential privacy can in some cases give better bounds than simply summing the privacy loss parameters.)

In a future version of \thesystem\ different tiers of access will be defined for data analysts. In the case of semi-trusted data analysts (who we assume will not collude, as discussed in Section~\ref{sec:actors}), \thesystem\ will provide {\em each} analyst a {\em per-user} privacy budget of $\epsilon_a$.   

In the case of completely untrusted analysts, we will {\em share} $\epsilon_a$ among all future analysts. The incorporation of these tiers of access into the system is under development. The latter model is more conservative with respect to privacy protection, and thus may be appropriate when analysts do not have the sufficient accountability or the data is highly sensitive (e.g.~with life-or-death or criminal implications).   The downside of the more conservative model is that  it is vulnerable to a denial-of-service attack, where the first few data analysts, intentionally or inadvertently, deplete the entire privacy budget, leaving future analysts unable to make any queries.   This can be partly mitigated by rate-limiting the use of the privacy budget and by sharing all statistics computed publicly. It is also possible to reserve part of the privacy budget for untrusted analysts and part for trusted analysts, with each part being treated as described above.

\item[Budgeting for Interactive and Adaptive Queries:]

An additional subtlety in privacy budgeting comes from the fact that data analysts may choose their privacy loss parameters $(\epsilon_i,\delta_i)$ {\em adaptively}, depending on the results of previous queries.  In such a case, it is natural to try to use composition theorems as {\em privacy filters}~\cite{RogersRoUlVa16} --- for example, the $k$'th query would be allowed only if its privacy loss parameters $(\epsilon_k,\delta_k)$ do not cause the inequality of the Optimal Composition Theorem (Thm.~\ref{thm:optcomp}) to be violated.  Unfortunately, as shown in \cite{RogersRoUlVa16}, this strategy does not in general yield $(\epsilon_g,\delta_g)$ differential privacy overall.   However, more restrictive bounds, such as Basic Composition (Thm.~\ref{thm:basiccomp}), do yield valid privacy filters.  Consequently, for interactive queries for data analysts in \thesystem, the Optimal Composition Theorem (and its approximations) is only used within (non-adaptive) {\em batches} of queries; to compose across different batches, we use Basic Composition.
\end{description}

\section{Differentially Private Algorithms}
\label{sec:statistics}
\begin{description}[leftmargin=*,itemindent=.5em, listparindent=\parindent]
\item[Choice of Statistical Procedures:]  
While \thesystem\ is designed to be easily extensible so as to incorporate new algorithms from the rapidly expanding literature, the initial set of differentially private algorithms in \thesystem\ were chosen to support the most necessary statistics that are needed to provide immediate utility for social science research and data exploration.   Specifically, we include:
\begin{itemize}[leftmargin=*,itemindent=1em]
\item Univariate descriptive statistics, such as means, quantiles, histograms, and approximate cumulative distribution functions.   
From some of these, post-processing can also provide additional descriptive statistics at no additional privacy cost.
\item Basic statistical estimators, for inference about the population from which a dataset was sampled.   We have selected some of the most widely used statistical inference procedures in social science, such as matching algorithms and difference-of-means tests for causal inference, 
and low-dimensional linear, logit, probit and poisson regression. 

\item Per-row transformations for creating new features (variables) out of
  combinations of already existing ones.  These allow the previously
  described procedures to be leveraged to do more sophisticated
  computations on a broader range of questions~\footnote{For example, the (empirical) covariance between two attributes can be estimated by estimating the mean of a new attribute that is the product of the two original attributes (as well as the means of the original attributes), or the mean of a variable in a subpopulation can be computed from the mean of the product of that variable with a binary indicator for the subpopulation of interest, and the mean of the indicator.}.
\end{itemize}

We have chosen to initially implement differentially private versions of statistical methods that are widely used in social science\footnote{For example Krueger and Lewis-Beck \cite{krueger2008ols} in a survey of all 1796 quantitative articles published between 1990 and 2005 in three leading political science journals find that 30\% use simple linear regression, and a further 25\% use either Logit or Probit regression.  Similar studies show regression is used in 18\% of published articles in psychology, 25\% in the education research \cite{troncoso2010statistical} and over 40\% of articles in the New England Journal of Medicine \cite{sato15}, while in Public Health, 20\% use regression and 43\% use Logit or Probit models \cite{karran2015statistical}.} and where the differentially private algorithms give good performance at sample sizes we found in social science research. For our evaluation, that we will report in Section~\ref{sec:evaluation}, we have examined 80 such datasets from published works that used methods available in our differentially private library. 

These choices are also motivated in part by the data exploration tools that \thesystem\ will integrate with, and which we expect our data analysts to use. In particular, the \tworavens\ graphical data exploration tool (\url{http://2ra.vn}) provides descriptive statistics for each variable in a dataset, as well as graphical illustrations of its empirical distribution (e.g. a histogram or a probability density function) \cite{DOrazio16}.  \thesystem\ replaces these with the differentially private descriptive statistics it computes.

Per-row transformations allow for building more sophisticated analysis. Indeed, these transformations and univariate means are sufficient to express all the \emph{statistical queries} in the sense of~\cite{Kearns93}. In order to allow only transformations that are \emph{safe}, 
\thesystem\ allows only transformations that are per-row and that come from a restricted domain-specific language. 
This language also allows for either specifying or automatically inferring ranges for transformed variables from those of the original variables, so that these can still be enforced for the transformed variables. A more comprehensive discussion of the variable transformations is in Section \ref{sec:security}.

\item[Measuring Accuracy:]
The choice of accuracy measure, and how to represent it to users, is important both in the privacy budgeting tool as well as for data exploration by data analysts, who need to know how to interpret the noisy statistics provided by differential privacy.
For descriptive statistics, we have determined that 95\% confidence intervals are the simplest and most intuitive way to represent the noise introduced by differential privacy.
 
For many of the basic differentially private algorithms for descriptive statistics (such as the Laplace mechanism~\cite{DMNS06}), a theoretical worst-case analysis is also indicative of typical performance, so we use this to calculate the \emph{a priori} privacy-accuracy translation needed in the privacy budgeting tool.

For statistical inference procedures, the accuracy (e.g.\ size of a confidence interval obtained) is necessarily data-dependent, even without privacy.  (For example, using a $t$-test for mean estimation gives a confidence interval of size that depends on the empirical variance of the data.)    When incorporating such methods, \thesystem\ uses {\em conservative} confidence intervals, meaning that it ensures that the differentially private confidence interval includes the true value with probability {\em at least} .95.  Intuitively, we account for the noise introduced by differential privacy by making the confidence intervals larger ---  this ensures that analysts do not draw incorrect conclusions from the differentially private statistics (but more analyses may come out inconclusive, as we explain to users of the system).  To provide the \emph{a priori} accuracy bounds needed by the privacy budgeting tool, we intend to use ``rules of thumb'' based on experimental evaluation given $n$, $\epsilon$, the number of variables, and other available metadata.
\end{description}

\section{Software Architecture}
\label{sec:architecture}
We have implemented a prototype of \thesystem that is ready for preliminary deployment in a data repository.  As mentioned throughout the text, some features are still under development. In this section we will describe the current implementation. 

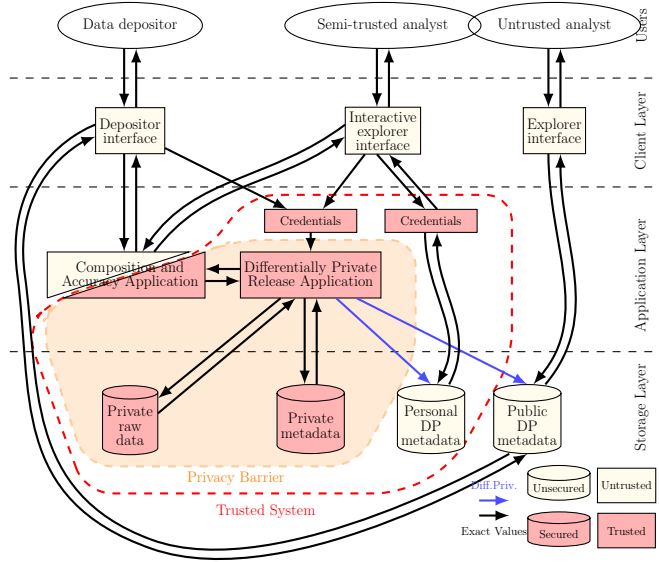
\begin{figure}[htb!]
\centering
\tikzstyle{block} = [rectangle, draw, fill=blue!20, 
    text width=5em, text centered, rounded corners, minimum height=6em, node distance=4cm,]
\tikzstyle{usercloud} = [ellipse, draw, node distance=3.5cm, 
	minimum height=4em]
\tikzstyle{prdb} = [cylinder, cylinder uses custom fill, 
	cylinder body fill=red!30, cylinder end fill=red!30,
 	shape border rotate=90, aspect=.25, minimum height=6em, draw, 
    node distance=5cm, minimum width=4.8em]
\tikzstyle{prdblegend} = [cylinder, cylinder uses custom fill, 
	cylinder body fill=red!30, cylinder end fill=red!30,
 	shape border rotate=90, aspect=.25, minimum height=3em, minimum width = 5.4em, draw, 
    node distance=5cm]

\tikzstyle{ptdb} = [cylinder, cylinder uses custom fill, 
	cylinder body fill=corn!50, cylinder end fill=corn!50,
 	shape border rotate=90, aspect=.25, minimum height=6em, draw, 
    node distance=5cm, minimum width=4.8em]
    
\tikzstyle{ptdblegend} = [cylinder, cylinder uses custom fill, 
	cylinder body fill=corn!50, cylinder end fill=corn!50,
 	shape border rotate=90, aspect=.25, minimum height=3em, minimum width = 5.4em, draw, 
    node distance=5cm]
\tikzstyle{localdb}= [rectangle, fill=corn!50, draw, text centered, minimum height = 4em, node distance=3.5cm]

\tikzstyle{localapplegend}= [rectangle, fill=corn!50, draw, text centered, minimum height = 3em, minimum width = 5em]
\tikzstyle{prlocaldb}= [rectangle, fill=red!30, draw, text centered, minimum height = 4em, minimum width = 2em, node distance=5cm]
\tikzstyle{cred}= [rectangle, fill=red!30, draw, text centered, minimum height = 2em, minimum width = 8em, node distance=5cm]

\tikzstyle{prlocalapplegend}= [rectangle, fill=red!30, draw, text centered, minimum height = 3em, minimum width = 5em]

\tikzstyle{semitrusted}= [ text centered, minimum height = 4em, minimum width = 13em, node distance=4cm]

\tikzstyle{line} = [draw, -latex']

\begin{tikzpicture}[node distance = 0.5cm, auto, scale=0.4, every node/.style={scale=0.4}]

\node [usercloud, align=center] (depositor) {\Large{Data depositor}};

\node [localdb, below of=depositor, align=center] (budgeter) {\Large{Depositor}\\\Large{interface}};

\node [semitrusted, below of=budgeter, yshift=-0.8cm, align=center] (c1) {};

\coordinate (A) at ($(c1) - (6.5em, 2em)$);
\coordinate (B) at ($(c1) - (-6.5em, 2em)$);
\coordinate (C) at ($(c1) + (6.5em, 2em)$);
\coordinate (D) at ($(c1) + (-6.5em, 2em)$);

\draw[red, thick, dashed, rounded corners=10, opacity=1] ($(A) + (1,-6.5)$) -- ($(A) + (-1,-1)$) -- (A) -- (C) -- ($(C) + (1.5,1.9)$) -- ($(C) + (10.2,1.9)$) -- ($(B) + (10.2,-2)$) -- ($(A) + (13.25,-5.5)$) -- ($(A) + (11.5,-6.5)$) -- cycle;

\node at ($(A) + (7,-7.1)$) {{{\color{red} \Large{Trusted  System}}}};

\draw[orange, thick, dashed, rounded corners=10, fill=orange!40, opacity=0.4] ($(A) + (1,-5.6)$) -- ($(A) + (-0.5,-1)$) -- (A) -- (C) -- ($(C) + (1,0.4)$) -- ($(C) + (7,0.4)$) -- ($(B) + (7,-2)$) -- ($(A) + (10,-5.6)$) -- cycle;

\node at ($(A) + (6,-6)$) {{{\color{orange} \Large{Privacy Barrier}}}};

  \draw[fill=red!30] ($(A) +(0.7em,0)$) -- (B) -- (C)  -- ($(A) +(0.7em,0)$);
  \draw[fill=corn!50] ($(A) -(0.7em,0)$) -- ($(C) -(1.4em,0)$) -- ($(D) -(0.7em,0)$)  -- ($(A) -(0.7em,0)$);

\node [semitrusted, below of=budgeter, yshift=-0.8cm, align=center] (compositor) {\Large{Composition and}\\ \Large{Accuracy Application}};

\node [prlocaldb, right of= compositor,xshift=1cm, align=center] (releaser) {\Large{Differentially Private}\\ \Large{Release Application}};

\node [cred, above of= releaser, yshift=-3.2cm, align=center] (credentials) {\large{Credentials}};

\node [prdb, below of=compositor, align=center] (privatedb) {\Large{Private}\\ \Large{raw}\\ \Large{data}};
\node [prdb, below of=releaser, align=center] (privatemeta) {\Large{Private}\\ \Large{metadata}};
\node [ptdb, right of=privatemeta,xshift=2.2cm, align=center] (dpmeta) {\Large{Public} \\ \Large{DP}\\ \Large{metadata}};
\node [ptdb, right of=privatemeta, xshift=-1cm, align=center] (dpmeta2) {\Large{Personal}\\ \Large{DP}\\ \Large{metadata}};

\node [usercloud] at ($(dpmeta2 |- depositor) + (-1.6,0)$)  (trusted) {\Large{Semi-trusted analyst}};

\node [usercloud] at ($(dpmeta2 |- depositor) + (4.1,0)$)  (analyst) {\Large{Untrusted analyst}};

\node [localdb, below of=trusted, align=center] (explorer) {\Large{Interactive}\\ \Large{explorer}\\ \Large{interface}};

\node [localdb, below of=analyst, align=center] (explorer2) {\Large{Explorer}\\ \Large{interface}};

\node [cred] at (dpmeta2 |- credentials)  (credentials2) {\large{Credentials}};

\draw ($(depositor.south)+(-0.2,0)$) edge[->, thick] ($(budgeter.north)+(-0.2,0)$);
\draw ($(budgeter.north)+(0.2,0)$) edge[->, thick] ($(depositor.south)+(0.2,0)$);

\draw ($(analyst.south)+(-0.2,0)$) edge[->, thick] ($(explorer2.north)+(-0.2,0)$);
\draw ($(explorer2.north)+(0.2,0)$) edge[->, thick] ($(analyst.south)+(0.2,0)$);

\draw ($(trusted.south)+(-0.2,0)$) edge[->, thick] ($(explorer.north)+(-0.2,0)$);
\draw ($(explorer.north)+(0.2,0)$) edge[->, thick] ($(trusted.south)+(0.2,0)$);
\draw ($(explorer.south)+(-0.6,0)$) edge[->, thick] ($(credentials.north)+(0.4,0)$);

\draw ($(compositor.north)+(0.8,0)$) edge[in=215, out=53, ->, thick] ($(explorer.west)+(0,-0.2)$);
\draw ($(explorer.west)+(0,0.2)$) edge[in=55, out=217, ->, thick] ($(compositor.north)+(0.4,0)$);

\draw ($(budgeter.south)+(-0.2,0)$) edge[->, thick] ($(compositor.north)+(-0.2,0)$);
\draw ($(compositor.north)+(0.2,0)$) edge[->, thick] ($(budgeter.south)+(0.2,0)$);

\draw ($(releaser.west)+(0,0.2)$) edge[->, thick] ($(compositor.east)+(0,0.2)$);
\draw ($(compositor.east)+(0,-0.2)$) edge[->, thick] ($(releaser.west)+(0,-0.2)$);

\draw [->, thick] plot [smooth, tension=0.95] coordinates { ($(dpmeta.south)+(-0.6,0)$)  ($(privatedb) + (0.7,-3.8)$) ($(compositor) + (-3.5,0)$) ($(budgeter.west)+(0,-0.2)$)};

\draw [->, thick] plot [smooth, tension=1] coordinates { ($(budgeter.west)+(0,0.2)$) ($(compositor) + (-3.75,0)$) ($(privatedb) + (0.7,-4.1)$) ($(dpmeta.south)+(0,0)$)};

\draw (budgeter) edge[->, thick] (credentials);
\draw (credentials) edge[->, thick] (releaser);

\coordinate (div0) at ($ 0.5*(budgeter)+0.5*(depositor)$);
\coordinate (div1) at ($ 0.61*(budgeter)+0.39*(compositor)$);
\coordinate (div2) at ($ 0.25*(budgeter)+0.75*(privatedb)$);
\coordinate (query3) at ($(privatedb)+(-5,3)$);

\draw [dashed] ($(div0) + (-4,0) $) -- ($(div0) + (17.5,0) $);
\draw [dashed] ($(div1) + (-4,0) $) -- ($(div1) + (17.5,0) $);
\draw [dashed] ($(div2) + (-4,0) $) -- ($(div2) + (17.5,0) $);

\draw ($(releaser.south)+(-1,0)$) edge[->, thick] ($(privatedb.east)+(0,0.9)$);
\draw ($(privatedb.east)+(0,0.4)$) edge[->, thick] ($(releaser.south)+(-0.5,0)$);

\draw (releaser) [blue!70] edge[->, thick] (dpmeta.north);
\draw (releaser) [blue!70] edge[->, thick] (dpmeta2.north);

\draw ($(releaser.south)+(-0.2,0)$) edge[->, thick] ($(privatemeta.north)+(-0.2,0)$);
\draw ($(privatemeta.north)+(0.2,0)$) edge[->, thick] ($(releaser.south)+(0.2,0)$);

\draw ($(explorer2.south)+(-0.2,0)$) edge[in=50, out=270, ->, thick]  ($(dpmeta.north)+(0.2,0)$);
\draw ($(dpmeta.north)+(0.6,0)$) edge[in=270, out=50, ->, thick] ($(explorer2.south)+(0.2,0)$);

\draw ($(credentials2.south)+(-0.2,0)$) edge[in=62, out=270, ->, thick] ($(dpmeta2.north)+(0.2,0)$);
\draw ($(dpmeta2.north)+(0.6,0)$) edge[in=270, out=62, ->, thick] ($(credentials2.south)+(0.2,0)$);

\draw ($(explorer.south)+(-0.2,0)$) edge[->, thick] ($(credentials2.north)+(-0.2,0)$);
\draw ($(credentials2.north)+(0.2,0)$) edge[->, thick] ($(explorer.south)+(0.2,0)$);


\node at ($(query3)+(17.1,-5)$) [blue!70] {Diff.Priv.};
\node at ($(query3)+(17.1,-6.5)$) {Exact Values};
\draw ($(query3)+(16.6,-5.5)$) edge[->, thick, blue!70] ($(query3)+(17.6,-5.5)$);
\draw ($(query3)+(16.6,-6)$) edge[->, thick] ($(query3)+(17.6,-6)$);

\node at ($(query3)+(21.5,-5)$) [localapplegend] {Untrusted};
\node at ($(query3)+(21.5,-6.5)$) [prlocalapplegend] {Trusted};
\node at ($(query3)+(19.25,-5.1)$) [ptdblegend] {Unsecured};
\node at ($(query3)+(19.25,-6.6)$) [prdblegend] {Secured};

\node at ($(depositor) + (17,0)$) {\rotatebox{90}{\Large{Users}}};
\node at ($(budgeter) + (17,0)$) {\rotatebox{90}{\Large{Client Layer}}};
\node at ($(compositor) + (17,0)$) {\rotatebox{90}{\Large{Application Layer}}};
\node at ($(privatedb) + (17,0.6)$) {\rotatebox{90}{\Large{Storage Layer}}};

\end{tikzpicture}\\
\caption{\em Architecture diagram.}
\label{fig:arch}
\end{figure}

\begin{description}[leftmargin=*,itemindent=.5em, listparindent=\parindent]
\item[Metadata:] Archival data for a research study are commonly stored on repositories as an original data file, and a complementary meta-data file.  The original data file contains the raw numeric values of observations in the dataset.
The meta-data file contains auxiliary information about the dataset that increases its ability to be reused by researchers; this might include text descriptions of the variables, summary statistics, provenance \cite{Cheney09} and numerical fingerprints for validation\cite{Altman09}.  The largest repositories have shared standards for how this meta-data file should be constructed \cite{Weibel98,Blank04}, so that catalogs of data can be built across repositories \cite{Mcneill07,Rice09}, and software utilities can be reused and deployed across different institutions \cite{Wilkinson16}.

Some of the information that gets recorded in the metadata we consider public, such as the names and text descriptions of the meanings of the variables and the sample size. Some of the metadata, such as variable-level summary statistics, contains private information, even if aggregated.  Thus if the dataset contains private information, we consider its metadata to also be a private file that could potentially leak information.  It is compliant with the shared standards, however, for metadata to have missing or empty fields, so we can construct a reduced version of the private metadata, that only contains public information.  To this we can add differentially private versions of certain summary statistics, and still distribute the metadata file for public use, so long as the total privacy loss after composition (see section \ref{sec:budgeting}) is below the appropriate global parameter.  We call this the \emph{public metadata}. 

The bottom of Figure \ref{fig:arch} shows the private raw data, its accompanying private metadata, and the public metadata, residing in a storage layer in our system.  Surrounding them are the application layer tools for differential privacy, which run on a remote server.
The differentially private algorithms, the accuracy estimates, and the budgeting coordinated by the composition theorem, each discussed previously, are all implemented in the R programming language, which is widely used in the statistics and quantitative social science communities \cite{Rcore16}. 
We describe how they interlink below, as we trace out user's interaction with the system.  We expect to distribute all of these routines as an R package for easy reuse within the R environment (independently of \dataverse\ and \tworavens). In addition to this code on the server, there are client layer interfaces (written as thin HTML Javascript GUI's) that allow different types of users to interact with the system, but no direct access to the raw data.  We now describe our different key users (the same as introduced in section \ref{sec:actors}), and how their respective interfaces interact with the larger system, in turn.

\item[Depositor Interaction:]
At the time of budgeting the \emph{depositor interface} or \emph{privacy budgeting interface}, as for example in Figure \ref{fig:interface}, allows the \emph{data depositor} to construct a list of statistics they would like to release on the dataset.  This interface has no direct access to either the data or computations on the data; whenever the page requires a new computation,\footnote{As when the metadata for a statistic is completed, or a statistic is deleted, or when an accuracy value, or any global parameter is edited.} it copies the contents of the current page to a remote application that uses differential privacy composition theorems to re-partition the privacy budget among the current set of statistics (by scaling all of the $\epsilon_i$'s by the largest multiplicative factor that stays within the global privacy budget), and recalculates the corresponding accuracies. This remote process then recomputes and returns an updated list of privacy loss parameters and accuracy estimates associated with each selected statistic.  The frontend interface then rewrites the summary table in the right-hand panel of Figure~\ref{fig:interface} with these newly provided values, and waits for more actions from the user until another round of computation is required.  The backend composition process is memoryless, in the sense that no past version of the page persists or is stored, but every request of the backend begins an entirely new set of budgeting and accuracy computations.  For this reason, the connection between the frontend and backend does not have to be persistent.

When the depositor has finalized the list of statistics she wishes to make available, together with their privacy loss parameters, a table containing the chosen statistics and their associated metadata and privacy loss parameters is then submitted to another separate remote release application that computes all the DP statistics requested. This release tool checks the composition of the request with a trusted version of the composition application, which means that code to this point does not have to be trusted, so long as the global $\epsilon$ can be verified.  This is detailed on Figure~\ref{fig:arch} as a split box, representing that there are two instances of the same code, one listening and replying to client requests, which does not have to be trusted, and another copy that has to be trusted, but only interacts with the backend, and has no web connection so is easier to protect. The release tool is the only process that has access to the raw data which sits in secure storage in the Dataverse repository.  The application that calculates the DP releases does not reply to the depositor interface.  The architecture diagram in Figure~\ref{fig:arch}, shows the directions of communication between every piece of the system and one can trace out from this that any path from the raw data to any data analyst (or even the data depositor), has to pass through the DP channel from this application to the release of a differentially private value written to a metadata file.

\item[Analyst Interaction:] 
The differentially private statistics that are generated are released in a file of metadata associated with the securely archived data.  Everything in this metadata file can be made available for public browsing. In Figure~\ref{fig:arch}, we show an \emph{untrusted public analyst} who does not need to prove any credentials, and is able to access the public metadata file with the differentially private releases.  The public analysts can use the public metadata file in whatever manner they prefer.  However, since all this information is written in the repository metadata standards, a difficult to read XML file, we provide an \emph{explorer interface} that presents the information in a more easily interpretable graphical form, using a modified version of the \tworavens~ software \cite{honaker2014statistical}, described in the next section. This is a statistical platform that allows users to explore data in repositories by means of their metadata, so is a good match for this application where only the metadata is available to the user.  

Once \thesystem\ is integrated with a data repository, we will provide another tier of access to \emph{semi-trusted} users.  These are users for which the depositor has granted a user-specific privacy budget $\epsilon_a$ from which they can generate additional differentially private releases, beyond those included in the public release.  We expect these users will have some distinct university or research affiliation which can be verified by credentials and agree to terms of use.\footnote{For example, \dataverse verifies members of certain universities by \emph{Shibboleth}~\cite{Morgan04} using the Security Assertion Markup Language protocol.}  Their explorer interface includes both the exploratory ability of the untrusted analyst interface, and the budgeting ability of the depositor interface.  Again, these users can construct a list of queries they would like to make, partitioning their personal $\epsilon_a$ budgets among them, by assistance of the composition application.  

When they have a batch of statistics whose accuracies they find useful, the statistics are submitted to the differentially private release function which checks that the composition of statistics meets the available budget using a trusted copy of the composition application.  This may occur in several adaptive rounds, as they learn things about the data that inform their further exploration, until their budget is exhausted.  Each semi-trusted user will have their own personal metadata file.  This will start out with the same information in the public metadata, but each release will add the additional differentially private releases that have been paid from that user's personal $\epsilon_a$ budget.  Only the semi-trusted user will have access to this metadata file, by means of their credentials, and specifically in the terms of use we are trusting they will accord to, they will have agreed not to share these values in collusion with other users (as discussed in the trust model in Section~\ref{sec:actors}).

\item[Security of the prototype:] In developing \thesystem\ we concentrated on design choices that maximize its usefulness for its potential user community.  Nevertheless, we addressed also several of the security and side-channel vulnerabilities that have been raised in the literature about implementations of differential privacy~\cite{Haeberlen11,MTSSC12}. Most of these concerns are mitigated by our design choices. For example, \thesystem\ only allows its users to select built-in differentially private data analyses and run variable transformations before them, rather than allowing arbitrary analyses that are then automatically verified or converted to satisfy DP.  This restriction comes naturally with our goal of allowing exploratory data analysis by users without programming expertise. More details on the vulnerabilities and how \thesystem\ addresses them are Section \ref{sec:security}.
\end{description}

\section{Exploration Interface}

\begin{figure}[htb]
\centering
\includegraphics[width=.7\textwidth]{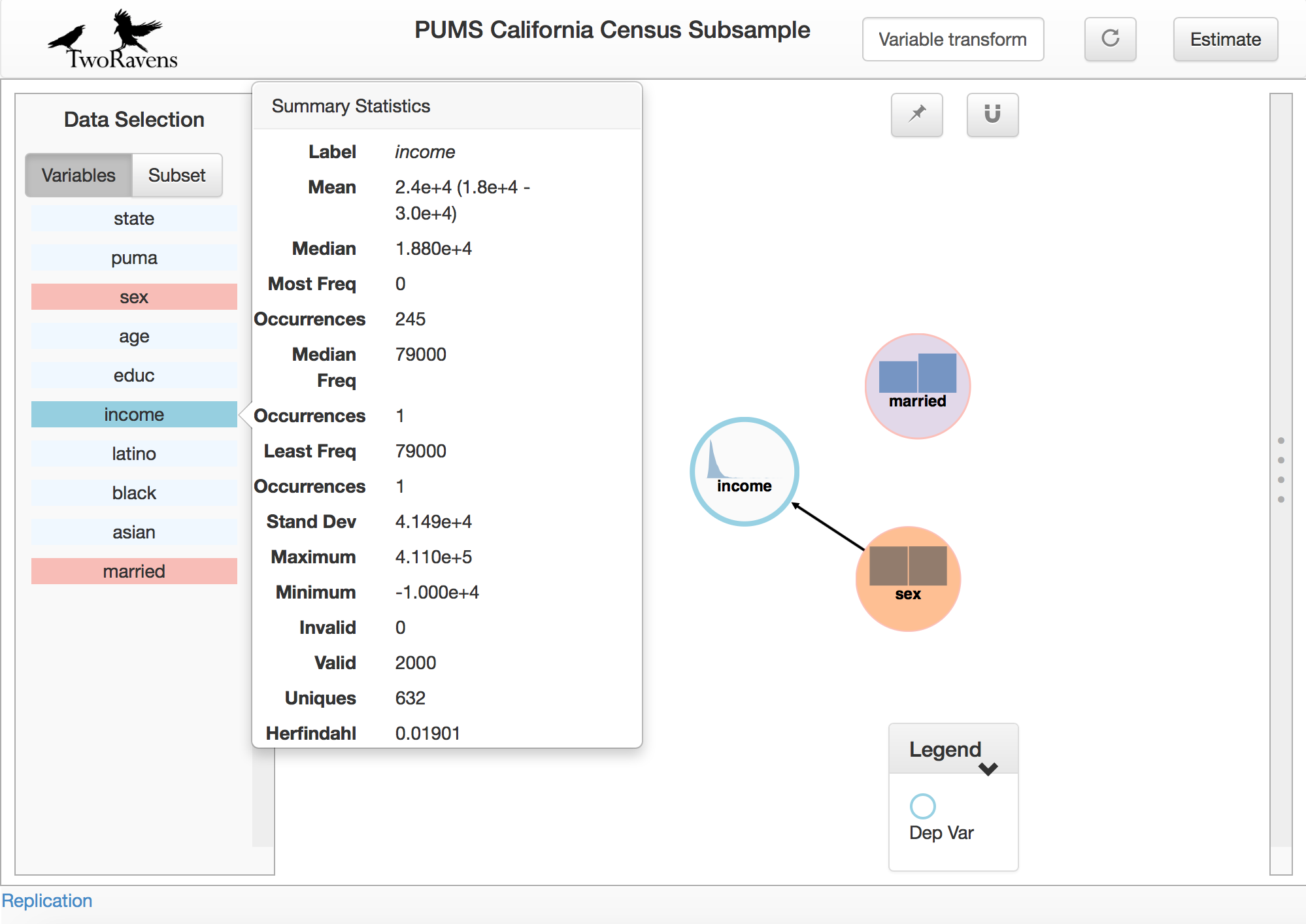}
\caption{\em Explorer graphical user interface for inspecting differentially private released values, adapting TwoRavens platform.}
\label{fig:explorer}
\end{figure}

As described in the previous section, all released differentially private values are written to metadata files, either public files or (in a future version of the system) files belonging to only one user.  These files can be used by the permitted analyst in whatever manner they prefer, but we provide in our system a user-friendly interface to read the information stored in the metadata.  The \tworavens\ platform for statistical inference (\url{http://2ra.vn}) is an interface that allows users, at all levels of statistical expertise, to browse data on repositories, explore summary statistics and build statistical models on those datasets by means of directed graphs \cite{honaker2014statistical, DOrazio16}.  The interface is a browser-based, thin client, with the data remaining in an online repository, and the statistical modeling occurring on a remote server.  The data remains in the repository and never goes to the browser; rather the statistical exploration is achieved by remote statistical processing and moving the correct metadata to the browser.  This architecture works well with the \thesystem\ system since it relies solely on metadata, and we have been adapting some of the graphs and summary tables available to convey to the user the additional uncertainty inherent in dealing with differentially private releases from noisy mechanisms, for example, providing confidence intervals for differentially private values, and histograms and density plots that represent the uncertainty in the values due to noise. 
\section{Security}
\label{sec:security}
The initial prototypes of \thesystem\ do not address all of the security and side-channel issues that have been raised in the literature about implementations of differential privacy~\cite{Haeberlen11,MTSSC12}.   We feel that a higher priority is evaluating whether the design of \thesystem\ is useful for its potential user community, and if the answer is positive, security issues can be addressed in a future version, before it is used to handle highly sensitive data.

\smallskip

\subsection{Timing, state and privacy budget attacks}

Haeberlen et al.~\cite{Haeberlen11} analyze the possible attacks to a differential
privacy system working in a centralized scenario similar to the one we
described in Section~\ref{sec:actors}. In their scenario, data analysts are
allowed to submit arbitrary analyses to the differential privacy
system and the system is responsible for running these analyses if they
pass some formal requirements guaranteeing differential privacy and if
there is still some budget left.  Even
if these formal requirements guarantee differential privacy, this
model is prone to three main kinds of side channel attacks:
\begin{description}
\item[Timing attacks] The data analysis may leak information
  about an individual using a timing (or any other covert) channel.
\item[State attacks] The data analysis may leak information about an
  individual through an observable change in the application state,
  for instance by using a global variable.
\item[Privacy budget attacks] The data analysis may leak information
  about an individual by running a subanalysis that fails because of
  lack of privacy budget.
\end{description}
Most of these attacks can be implemented only if data analysts are
allowed to submit arbitrary data analyses. In \thesystem a data
analyst can only select built-in differentially private data analysis and run variable
transformations before them. Using only built-in differentially
private data analysis prevents these attacks at data analysis
time.  
For instance, there is no risk of a privacy budget attack since
queries cannot run subanalyses that can exhaust the privacy budget. 
Nevertheless, data analysts can submit to \thesystem\
variable transformations that can create new features by combining
existing ones as we discussed in Section~\ref{sec:statistics}, and
these raise a greater risk of timing attacks and state attacks, which we
discuss in the next section. 
\begin{figure}[htb]
\centering
\includegraphics[scale=.30]{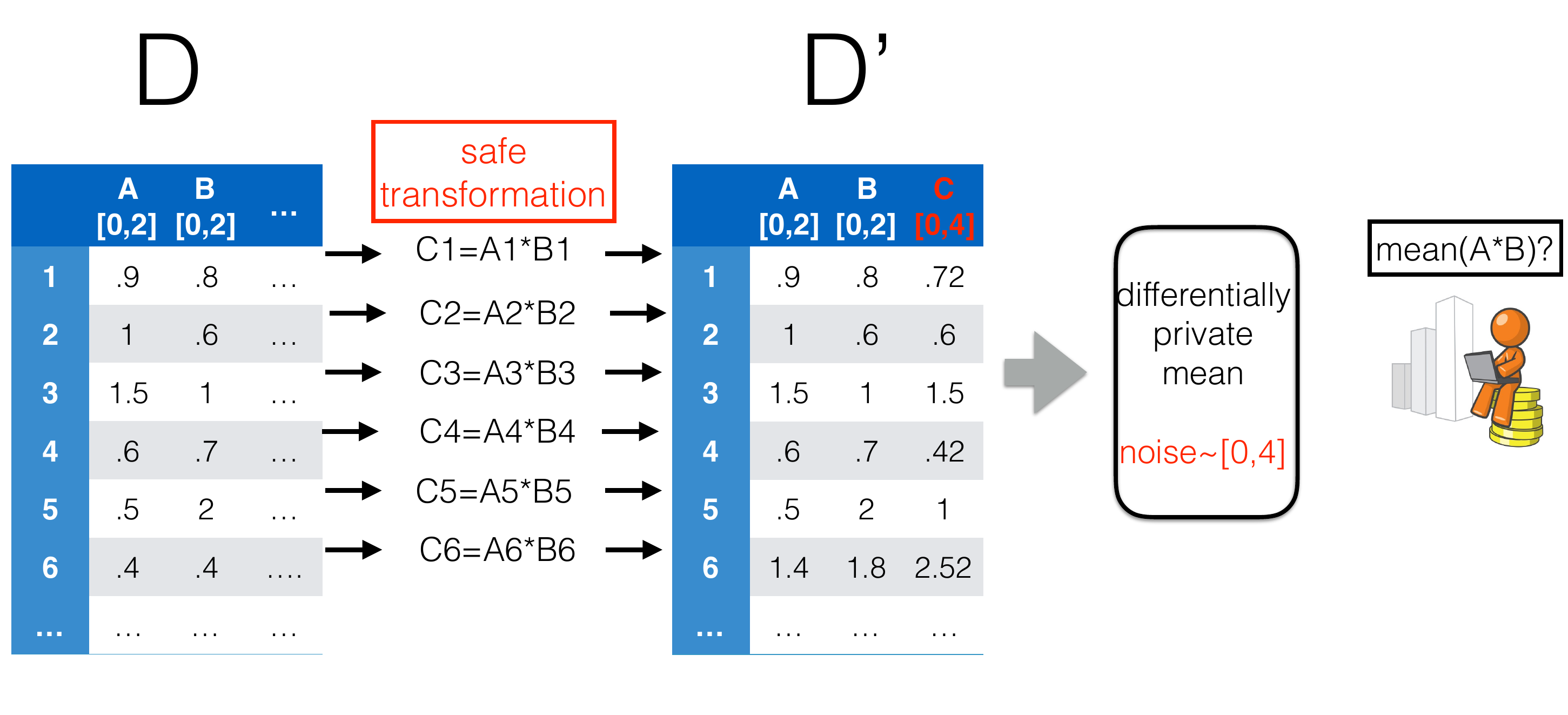}
\caption{\em Workflow schema for safe variable transformations.}
\label{fig:safeTransform}
\end{figure}

\subsubsection{Safe variable transformations}
An important property of differential privacy is closure under
post-processing.

\begin{lemma}[Post-processing, \cite{DMNS06}]
  Let $M$ be a $(\epsilon,\delta)$-differentially private randomized
  algorithm from\footnote{We use $X^n$ to describe the possible set of
    databases with $n$ records of type $X$.} $X^n$ to $Y$, and $f$ be an arbitrary (possibly randomized) map
  from $Y$ to $Z$. Then, the composition of $M$ and $f$, denoted
  $f\circ M$ is a $(\epsilon,\delta)$-differentially private
algorithm from $X^n$ to $Z$.
\end{lemma}

This says that the result of a differentially private
computation can be safely given as input to any other transformation and the
differentially privacy property will be maintained. 
The situation is more involved when 
transformations occur before applying a
differentially private mechanism. Indeed, if we fix a differentially private algorithm
$M$  from $X^n$ to $Y$ and we arbitrarily pre-process its input dataset $z\in Z^n$
with an arbitrary map $f$  from $Z^n$ to $X^n$ we can break its
privacy guarantee. 
As a simple example, consider a differentially private
mechanism that approximately releases the fraction of the individuals
with a particular feature $B$ in a database with $n$ records, and a map $f$ that returns a database with $n$ records with the feature
$B$ in the case John Doe is in the database, and that removes all the
elements with feature $B$, otherwise. When $n$ is sufficiently large, a
data analyst observing the result of $M\circ f$ can determine with
high probability whether John Doe is in the database or
not. 
Fortunately, there is a class of important transformations that
preserves differential privacy: per-row transformations. 

\begin{lemma}[Per-row transformations]
\label{lem:per-variable}
  Let $M(x_1,\ldots,x_n)$ be a $(\epsilon,\delta)$-differentially private randomized
  algorithm\footnote{We make here explicit the fact that $M$ is a
    function of the records $x_1, \ldots,
    x_n$ of the input dataset.} from $X^n$ to $Y$, and $f$ be a map
  from $Z$ to $X$. Then, the composition of $f$ and $M$, denoted
  $$(M\circ f)(z_1,\ldots,z_n)=M(f(z_1),\ldots, f(z_n))$$ is a $(\epsilon,\delta)$-differentially private
  randomized algorithm from $Z^n$ to $Y$.
\end{lemma}

This kind of transformation can be very useful in practice.
For instance, as we mentioned before, the (empirical) covariance between two attributes can be estimated by estimating the mean of a new attribute that is the product of the two original attributes (as well as the means of the original attributes), or the mean of a variable in a subpopulation can be computed from the mean of the product of that variable with a binary indicator for the subpopulation of interest, and the mean of the indicator.

However, one must be careful in using this lemma. Indeed, if the input
data is $Z^n$ one must consider the possible change of attributes in
$X^n$ when reasoning about the  differentially private algorithm
$M$. 
Let see this with an example. Consider the case where we want a differentially private estimate
of the mean of a new attribute $C$ that is the
product of the two original attributes $A,B$. A differentially
private algorithm for computing the mean must choose noise that is
proportional to the range of the attribute. Suppose that we
know the range of $A$ and $B$ is $[a,b]$ for $a,b\geq 0$. When we choose the noise
for $M$ we need to reason about the range of $C$ which is not $[a,b]$
but it is $[a^2,b^2]$ instead.

In order to allow only transformations that are \emph{safe} in the
sense discussed above,
\thesystem\ takes an approach similar to the one of AIRAVAT~\cite{RSKSW10} and requires the data curators and the data analysts to 
provide the ranges of each variable before and after the
transformations and enforces them at runtime, i.e. the differentially
private algorithms, truncate values that are outside
the specified range. This guarantees the correct use
of the principle formalized in Lemma~\ref{lem:per-variable} and so privacy is preserved. 

To support the design of transformations \thesystem\
uses a restricted domain-specific language and an automated
program analysis tracking variable ranges. 
The workflow
of variable transformations is described in Figure~\ref{fig:safeTransform}. Starting
from the private dataset $D$, a variable transformation generates a
new private dataset $D'$, containing the same individuals as $D$ but
with potentially new variables, on which the differentially private
algorithm is run. In this example, the transformation creates a new
attribute $C$ as the product of $A$ and $B$. This is performed per-row
and the program analysis forwards the information about the range from
the inputs (in this example the range for both $A$ and $B$ is $[0,2]$) to
the newly generated variable (in this example the range for $C$ is
then $[0,4]$). This range is provided to the user who can decide to
keep it or to use a different range. The differentially private
algorithm will then enforce this range and add noise proportional to it.

The language for variable transformations allows only
statistical operations that combine, transform or separate
variables in a value independent way. This prevents high-level
timing attacks --- ones where the timing leakage is intentional ---
even if it doesn't prevent fine-grained timing analysis on numerical
computations, as we will discuss below.
Moreover, to protect against state attacks, the language for variable
transformations only allows access to locally defined
variables. The program analysis is based on a flow-sensitive type system that is used to guarantee that information about the changes in the ranges of variables are propagated to the output. 

 Summing up, our approach of separating variable transformations from the
differentially private data analysis (whose code is not accessible by
the data analyst) guarantees protection against privacy budget
attacks. The use of a domain specific language further protects against
state attacks and (high-level) timing attacks. Finally, the enforcement
of the variable ranges at runtime prevents the misuse of the variable
transformations. To help the user decide the range for each
variable, the domain specific language uses a program analysis
propagating range information from the input to the output. 

\smallskip

\subsubsection{Floating-point rounding attack}
Another attack is the
floating-point rounding attack identified by
Mironov~\cite{Mironov12ccs}.  The idea of this attack is to exploit
the irregularities in floating-point implementations of some basic
algorithms like the Laplace mechanism. When the output is numeric,
differential privacy requires every output to be feasible, i.e. being
returned with some probability for every
input, and outputs to have similar probabilities when the inputs
differ by an individual.  Mironov showed instead that naive
implementations of differential privacy lead to results that are concentrated on subsets of outputs. Even worse, it can be the case that for neighboring databases some outputs may only be possible under one of the two databases. This allows adversaries to distinguish the output distributions of the two database with certainty and violate differential privacy. 

The solution proposed by Mironov is to use the snapping
mechanism~\cite{Mironov12ccs,DLAMV12} which essentially tosses out the least significant bits of the differentially private floating-point outputs using a combination of clamping and rounding procedures.  This
mechanism is also effective when the mechanism is instantiated with
imperfect randomness. We have implemented the snapping mechanism and incorporated it into the library of differentially private algorithms that underlies the system. However, we found that it has poor utility when compared to the typical Laplace mechanism so at this stage we do not offer it as a default through the interface. 

\smallskip

\subsubsection{Fine-grained side channels attacks}   
Side channel attacks are in general difficult to prevent. We discussed
before how the use of built-in differentially private primitives and a
domain specific language for variable transformations can help in
mitigating timing channels. Nevertheless, the current implementation
may still be prone to fine grained attacks like the one by~\cite{AndryscoKMJLS15}
exploiting time leakages due to floating points computations. We
expect these kinds of attacks to be further mitigated by the fact that
\thesystem\ is only accessed remotely and so some of these fine grained
observations are absorbed by delays in the communication.

We expect
that by using an execution environment where statistical operations
have value-independent cost, which can
be achieved by padding thanks to the restricted setting, by using some
of the proposed mitigations~\cite{AndryscoKMJLS15}, and by having \thesystem\ only
accessed remotely we can prevent further vulnerabilities.  Nevertheless, we leave a complete
evaluation of these vulnerabilities to a future version of our
prototype.
\section{Usability testing}
\label{sec:uiux}
We conducted thorough usability testing of \thesystem , seeing 28 participants in total. UI testing was broken into three phases: a pre-pilot phase ($n=3$), a pilot phase ($n=5$), and a full study ($n=20$). After each phase of testing, improvements were made to both the tool and the study protocols based on feedback from the participants. Because the study procedures differed in the three phases and the participants in the full study tested a more up to date version of the tool than was used in the pilot phases, we will primarily focus on the full study in this section. User testing was conducted on a pared down version of the system. 

We tested the workflow of data depositors releasing univariate statistics through the privacy budgeting interface. From a user perspective, the acts of releasing multivariate statistics and using the interactive query interface involve very similar procedures to the ones required for only releasing univariate statistics through the budgeting interface. For this reason, we focused on a core set of tasks that would shed light on the usability of the whole system while minimizing the fatigue of research participants. The tool for variable transformations discussed in Section \ref{sec:security} was not integrated into the system at the time of testing so has only undergone internal testing and will be evaluated in future rounds of user experiments as more features are added to the system.

All participants were over 18 years old and had some experience with data analysis, either through courses, work, or research. The study was approved by Harvard's Institutional Review Board and all participants were compensated with a $\$20$ gift card to Amazon for sessions lasting about one hour. In the full study, participants' education levels varied, with 10\% having completed some college, 25\% with a Bachelor's degree,  50\% with a Master's degree, and 15\% having attained a PhD. 35\% of participants reported being unfamiliar with differential privacy, 50\% said they were somewhat familiar, 10\% described themselves as familiar, 5\% were very familiar and nobody reported being an expert in differential privacy. 

\paragraph{Study procedures}
The user tests were designed to simulate the experience of data depositors. First, participants were asked to read brief introductory text in \thesystem\ broadly describing the purpose of the tool, the concept of differential privacy and privacy loss parameters, the need for metadata, and the idea behind secrecy of the sample. Next the participants were given a scenario designed to simulate the mindset of a data depositor. They were given a toy dataset in Excel containing the demographic information (age, sex, income, education level, race, and marital status) of 1000 people sampled randomly from a county with population 700,000. They were told that their goal was to advertise their dataset to other social scientists who were interested in the relationship between race and income for people of various ages. After reviewing the dataset, participants were asked to set privacy loss parameters for the scenario and were given a choice whether or not to use the secrecy of the sample feature. 

After setting privacy loss parameters, participants in the full study were led through a sequence of 11 tasks using the interface. The tasks all related to the scenario and toy dataset and required participants to effectively use each feature of the tool, guiding them through the typical workflow of a data depositor. 
The tasks are listed below with each of the features being tested indicated in parentheses. 
\begin{enumerate}
\itemsep0em 
\footnotesize{
    \item You just entered a tutorial mode in the interface that will highlight some key features of the tool. Go through the tutorial and, when prompted, select a mean of the Age variable as your first statistic. (Tutorial mode, selecting statistics, inputting metadata).
    \item You decide that the income and race variables are also important for future researchers, so you decide to release statistics for these. Add a mean and a quantile for income, as well as a histogram for race. (Selecting statistics, inputting metadata). 
	\item You no longer wish to include a quantile for income. Delete this statistic. (Deleting statistics). 
	\item	You decide that you want to be very confident in your error estimates. Use the tool to set a 98 percent confidence level. (Adjusting confidence level). 
	\item	You are thinking about your dataset, and you realize that it contains some information that makes it more sensitive than you originally thought. Use the tool to make the changes necessary to reflect this shift. (Adjusting privacy loss parameters). 
	\item	You have just been informed by a colleague that your dataset was actually randomly sampled from a population of size 1,200,000. Use the tool to make changes to reflect this. Does this make your statistics more or less accurate? (Secrecy of the sample). 
	\item	You decide that it would be useful to allow other researchers who do not have access to your raw data to make some of their own selections for statistics to calculate from your dataset. Use the tool to make changes to reflect this. (Reserving privacy budget for data analysts). 
	\item	How much error is there for the mean age statistic? What does this number mean? (Interpreting error estimates).
	\item	Make it so that the released mean age is off from its true mean by at most one year. Is this more or less accurate than what you had before? (Redistributing privacy budget via error estimates). 
	\item	Make it so that each count in the released race histogram is off from the true count by at most 5 people without changing the error you just set for mean age. (Redistributing privacy budget, hold feature). 
	\item	You are satisfied with your statistics and your error estimates. Finalize your selections. (Submitting statistics). 
	}
\end{enumerate}

In the two pilot phases, the tasks were much less specific and simply asked participants to use the tool to release some useful statistics about their dataset. Pilot participants reported feeling overwhelmed by the open-ended nature of this request so we modified the tasks in the full study to specifically target each feature of the tool. 

Audio recording was used during tasks and participants were asked to speak their thoughts out loud as much as possible, which provided valuable qualitative feedback. For each participant, time spent on each task was recorded. Errors were also recorded and classified as either Critical Errors (CEs), in which the participant made choices that led to the inability to complete the task, and  Non-critical Errors (NCEs), where participants corrected their own errors and were able to successfully complete the task despite early mistakes.  

\begin{figure}
\centering
\parbox{0.55\textwidth}{
\begin{tabular}{|c|c|c|c|}
\hline
Task & \begin{tabular}{c}Average time\\ on task (secs)\end{tabular}& \# CEs & \# NCEs \\
\hline
1 & 349.9 & 2 &1  \\
\hline
2 & 289.7 &9  &6  \\
\hline
3 & 8.2 &2  &1  \\
\hline
4 &  29.8 & 2 &  0 \\
\hline
5 & 53.7 & 4 & 2  \\
\hline
6 &  30.1 &1  &0  \\
\hline
7 &  95.8 &5  &1  \\
\hline
8 & 34.3 & 2 & 0  \\
\hline
9 &  20.3 &5  &3  \\
\hline
10 &  52.4 & 6 & 3  \\
\hline
11 & 20.6 &0  &0  \\
\hline
\end{tabular}
}\parbox{0.1\textwidth}{\includegraphics[trim={0 2cm 0 1.8cm},clip,width=2.5cm]{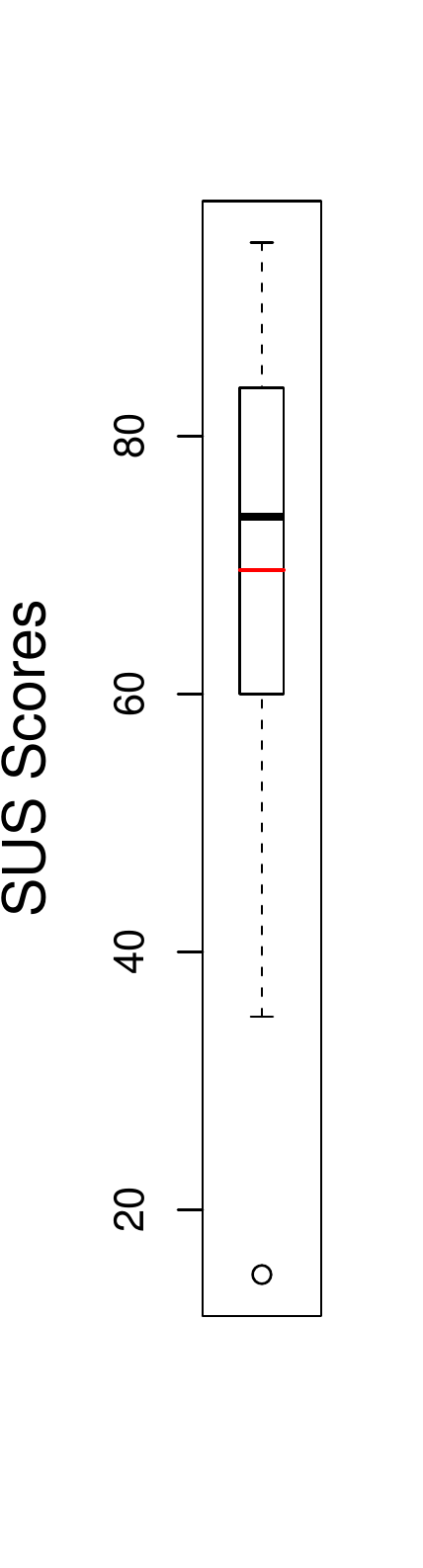}}
\caption{\textbf{Performance on usability tasks:} \it Average time on task is reported only for participants who successfully completed the task. The \# CEs and \# NCEs columns list the number of people who committed at least one critical error or non-critical error, respectively, during the task.  On the right is a box plot of System Usability Scale scores for \thesystem. The black vertical line is the median (73.6) and the red vertical line is the mean (69.6). The circle represents one outlier in the data.}
\label{tab:usability}
\end{figure}

\paragraph{Results}
Overall, participants performed well, committing relatively few critical errors during the tasks. The most common mistakes occurred when entering metadata, with every participant reporting some doubt over what values to enter and 45\% of participants entering values at some point in the session that would lead to poor results. 30\% of participants had some degree of trouble figuring out how to redistribute their privacy budget across their statistics (by raising or lowering the corresponding accuracy estimate).  The basic functionalities of selecting and deleting statistics, modifying global parameters, and submitting the statistics for differentially private release came easily to most participants. Mistakes tended to recur across subjects, clearly highlighting the more difficult features of the tool. A quantitative summary of the task results can be found in Table \ref{tab:usability}. Anecdotally, most participants said the documentation was helpful but should be simpler, shorter, and distributed as needed throughout the tool. Unsolicited, four participants reported having fun using the tool. 

There was no significant relationship between familiarity with differential privacy and the total number of critical or non-critical errors participants made. Likewise, there was no significant relationship between education level and the total number of critical or non-critical errors made. 

\thesystem puts some privacy-critical decisions into the hands of its users, making precise documentation, intuitive design, and thorough error-handling imperative. There are three ways that data depositors could accidentally leak more information about their data than intended: mistakenly setting overly large privacy loss parameters, overestimating the size of the population using the secrecy of the sample feature, or entering values directly from the raw data as metadata in the tool (note that none of these violations is possible for data analysts). Over all participants including the two pilot phases and all tasks, 0 people overestimated the population size when using the secrecy of the sample feature even though the majority of participants elected to use it. One participant in the pilot phase entered empirical range values from the toy dataset in the metadata field. In reaction to this, we added an additional warning about the risks of data-dependent decisions in the tool which was heeded by all 20 members of the full study. Only one participant of the 28 made an unsafe choice regarding the privacy loss parameters by accidentally switching $\epsilon$ and $\delta$, setting $\epsilon=10^{-6}$ and $\delta=.25$. Although a rare event, we have taken this error seriously and have implemented more rigorous checks on the privacy loss parameters to prevent similar accidents in the future.  All other participants followed the instructions in the documentation, setting $\epsilon$ to a small constant ($\epsilon$ values chosen by all other participants ranged from .05 to 1) and $\delta$ to a negligible number (range: $10^{-7}$ to $10^{-5}$). These results suggest that the system is designed in such a way that makes it difficult to violate privacy even for users who make many mistakes while using the tool.  

At the end of the usability test, participants were asked to rank how relevant they think \thesystem is for people who collect human-subjects data on a scale from 1 to 5, with 1 being the least relevant and 5 being the most. The average relevance rating given to the project was 4.3. We view this as an indication that people both recognize data privacy as an important issue and believe that \thesystem would successfully address a need in this space. 

\paragraph{System Usability Scale}
After the tasks, participants filled out the System Usability Scale (SUS) \cite{brooke96}, a ten item questionnaire that is widely employed in usability studies to assess the quality of a user interface. The SUS is easy to administer and though only ten questions, has been shown to be a reliable and valid measure of usability \cite{sauro11, bkm08}. The ten questions are on a five point Likert scale and yield a total score between 0 and 100, which should not be interpreted as a percentile or letter grade. 

In the full study, the mean SUS score given to PSI was 69.6 with a median score of 73.6. A box and whisker plot of SUS scores can be seen in Figure \ref{tab:usability}. There was one strong outlier in the data, more than 2.5 standard deviations below the mean. Removing this outlier gives a mean SUS score of 72.5 and a median of 75. These scores are better than the system's average score in the pilot study of 59.5, suggesting that changes made to the tool between the pilot and the full study made the system significantly more usable. There was no significant relationship between SUS scores and familiarity with differential privacy nor education level.

Two meta-analyses have been conducted on a wide range of usability studies and found average SUS scores across all systems to be 69.7 \cite{bkm08} and 68.0 \cite{sauro11}, respectively. There has also been work on associating labels with SUS scores \cite{bkm08} including: Worst Imaginable, Awful, Poor, OK, Good, Excellent, Best Imaginable. Of the participants who rated a system OK, the average SUS score was 52.0 and of those who selected Good, the average SUS score was 72.8. In a study on well known user interfaces, Excel received an average SUS score of 56.5, people gave their GPS's an average rating of 70.8, while the iPhone received a 78.5 average\cite{kb13}.  In light of these results and the complexity of \thesystem, we find our scores on the System Usability Scale encouraging. 

\paragraph{Incorporating Feedback}
After each phase of testing, modifications were made to the design of the tool and the documentation in accordance with user performance and feedback. A major change inspired by the usability test was the incorporation of a tutorial mode that is automatically triggered when users first encounter the interface. The tutorial orients users to the features of the tool and guides them through selecting their first statistic. There were also substantial changes made to the documentation with a focus on simpler, more concise language and more intuitive and visual locations of help text throughout the interface. Many more smaller changes were made to improve the usability of the UI, including new buttons, bug fixes, hiding advanced features when they're not needed, and a host of cosmetic adjustments. 
As mentioned in the results section, a particular effort was made after each testing phase to prevent accidental privacy leaks. The most significant of these was a more rigorous automatic checking system during privacy loss parameter selection. These checks work to prevent accidental unsafe parameter settings and provide clear alert messages if imprudent choices are made.  

\vspace{-1ex}

\section{Empirical Evaluation}
\label{sec:evaluation}
In addition to user testing, we have experimentally evaluated all of the differentially private algorithms implemented in \thesystem\ using a combination of real and synthetic data.   

We have performed two main kinds of experimental evaluations:
experiments aiming at confirming the feasibility of releasing several statistics with a given budget,
 experiments replicating studies from the social science literature.

\paragraph{Experiments on the combined release of statistics} 
The goal of this category of experiment was to answer the question ``can we release basic statistics for all the variables with a fixed budget and with a good accuracy?''

We have analyzed several datasets of different size (with $n$ as small as $10^3$ and as big as $10^6$) available in Dataverse.  The overall goal was to release all the univariate statistics currently implemented in \thesystem\ under different values of the budget for $\epsilon$ (in the range $[0.01,1]$) with fixed $\delta$ (set at $2^{-20}$) and varying the secrecy of the sample assumption (with values $1\%, 3\%, 5\%, 100\%$). We have considered different splits of the privacy loss parameters among the different statistics, and we have experimented using the optimal composition theorem and the basic composition theorem. We have also used different accuracy measures to capture different characteristics of the different data: mean absolute error, mean relative error, mean squared error, root of mean squared error, $\ell_1$, $\ell_2$ and $\ell_\infty$ norm. 

\begin{figure}[htb]
    \centering
   \includegraphics[clip=TRUE, height=4cm]{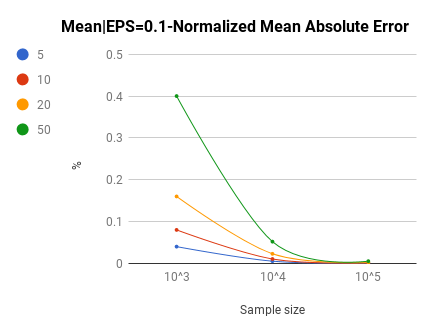}
   \includegraphics[clip=TRUE, height=4cm]{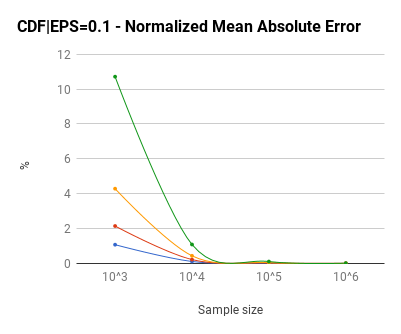}
   \includegraphics[clip=TRUE, height=4cm]{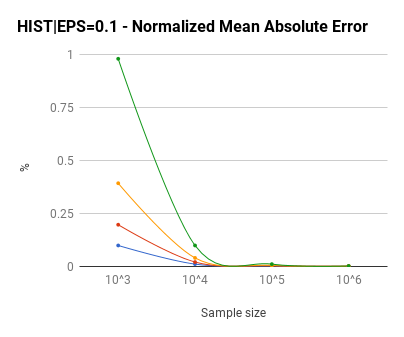}
    \caption{\it Normalized mean absolute error for  Means, Histograms, and CDFs for (5,10,20, and 50) variables from the PUMS dataset.}
\label{fig:budget-studies}
\end{figure}

From this experience we learned that in many situations we can provide differentially private results for all the univariate statistics with a non-trivial accuracy. For datasets with sample size 100,000 we were able to release several univariate statistics (mean, histograms, and CDF) for all the variables ($\sim$50 attributes), with mean absolute error $\leq 10\%$, with  global $\epsilon=0.3$ and global delta $2^{-20}$,  as shown in Figure~\ref{fig:budget-studies}.

As expected, these results have shown some variability depending on the setup of the parameters, e.g. larger dataset sizes and larger values of epsilon give better accuracy, as well as on the error metric used to measure accuracy. Nevertheless, the experiments we performed met some of the expectations set in Section~\ref{sec:incentives} and confirmed what the theoretical analysis tell us. Besides, 
this step helped us optimizing the code of the different statistics increasing the scalability of the analysis, e.g. the release of several univariate statistics (mean, histograms, and CDF) on datasets with milions of entries and $\sim$50 variables takes less than 10sec. 
\paragraph{Replication of social science studies}  
Replicating the results of published works is a higher bar for \thesystem\ than the actual initial goal, which is to support data exploration (for determining whether one should apply for access to raw data). Nevertheless, we created a corpus of 80 datasets from quantitative social science by finding datasets on repositories and reaching out to authors of studies.  Our goal was to find datasets that could be publicly released, but whose topics and structure closely resembled those that would ordinarily be closed due to the inclusion of sensitive data.  These give us a variety of types and sizes of datasets from which we can benchmark the performance of differentially private statistics, while releasing in comparison the true dataset values.  From this corpus, we also chose twelve studies which had published articles or reports using simple statistical methods that we could emulate using our available differentially private statistics.  These ranged in size from 926 observations in a survey of high school biology teachers and whether they teach evolution in the classroom to 369,000 observations in a randomized field experiment testing the ability to mobilize voters by pride or shame by mailing them their previous turnout history.  As an exemplar, we briefly describe one replication study.  

\begin{figure}[htb]
    \centering
   \includegraphics[trim={1cm 0 13cm 0}, clip=TRUE, height=5.5cm]{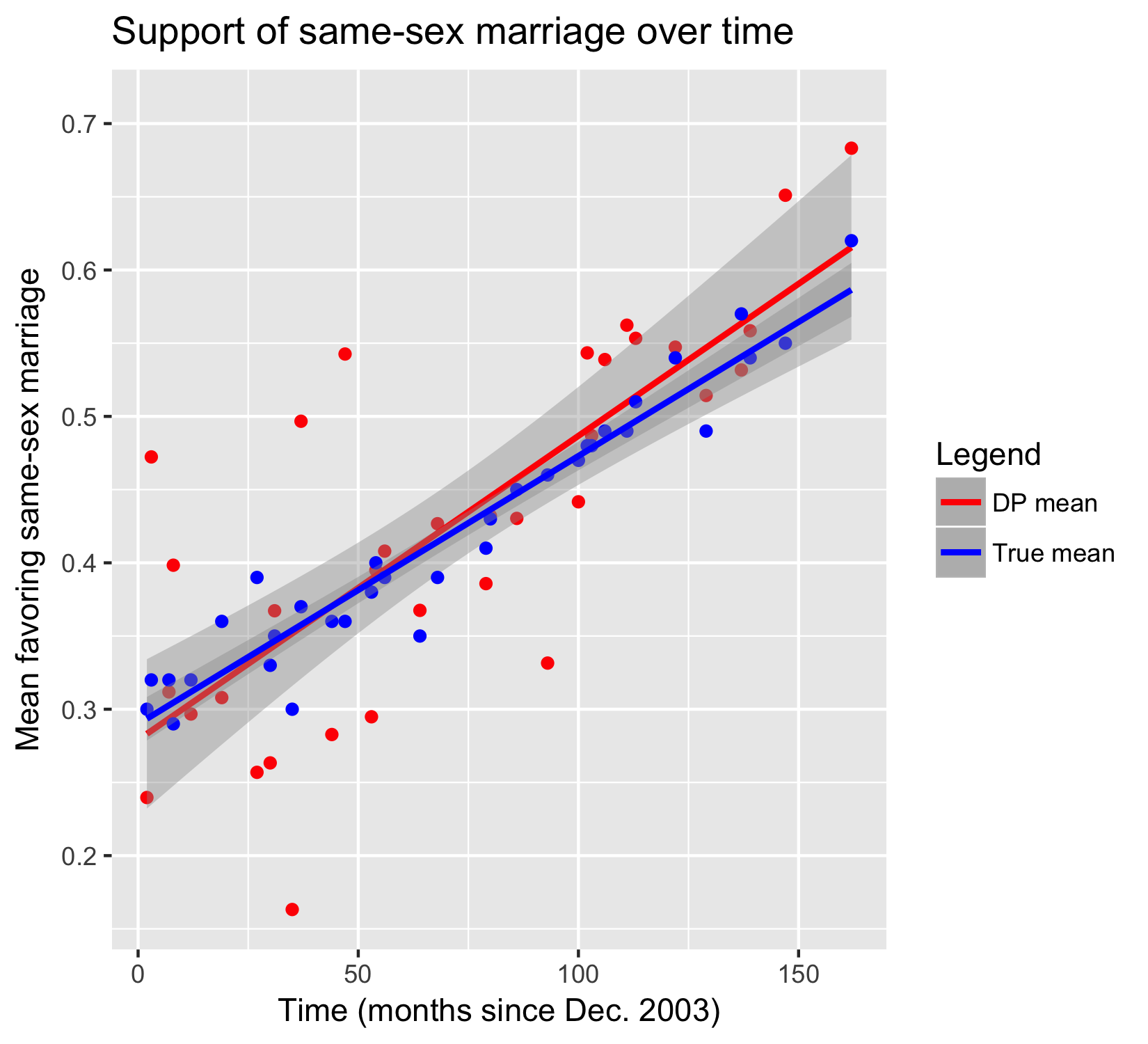}
   \includegraphics[trim={0 0 1.5cm 0}, clip=TRUE, height=5.5cm]{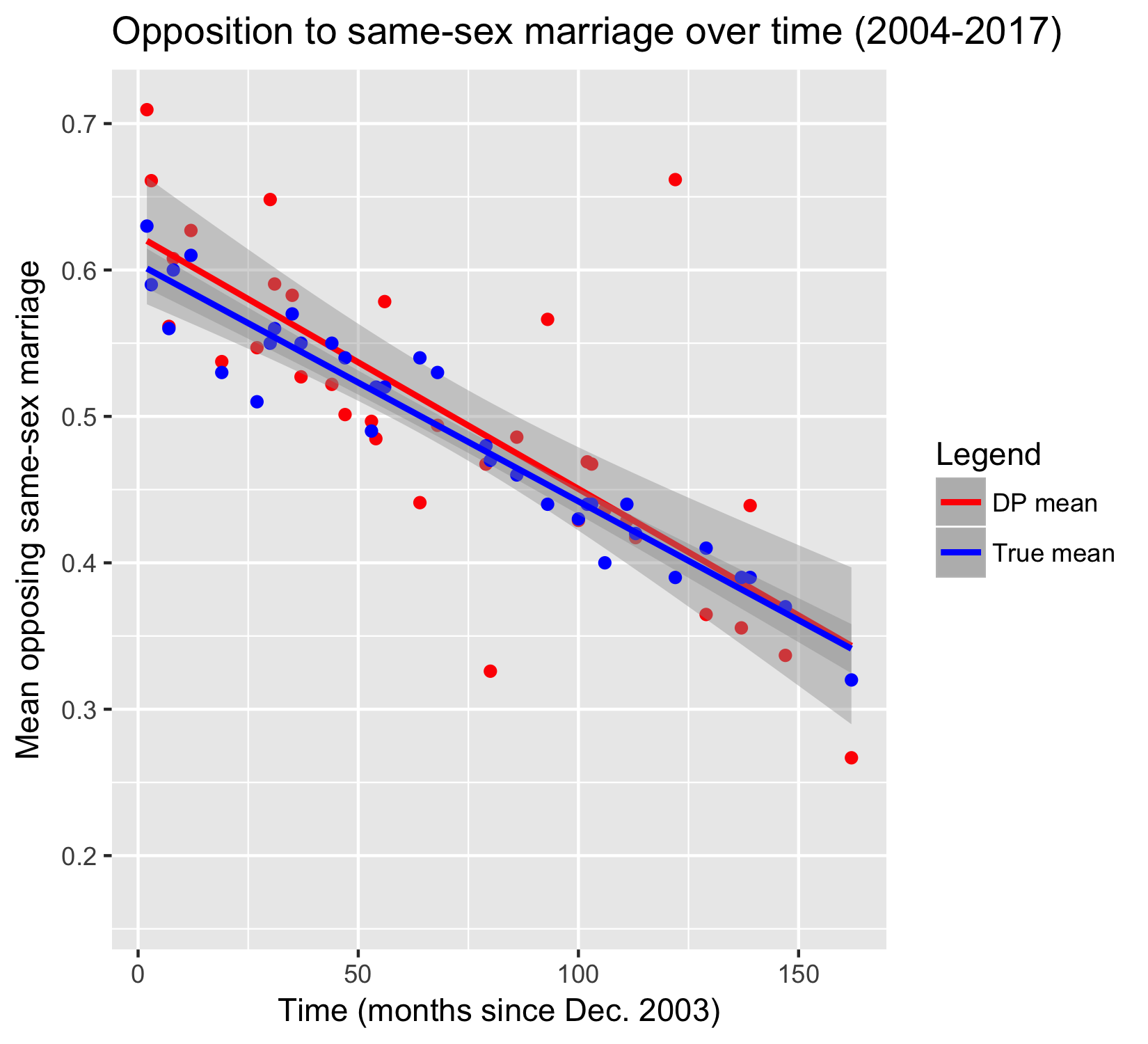}
    \caption{\it Pew opinion data over time, and estimated trends.}
\label{fig:pew}
\end{figure}

The Pew Research Center conducts periodic public opinion polling of attitudes and social and demographic factors.  They released a report describing trends over time in their data with regard to attitudes toward same-sex marriage \cite{Masci2017, PewResearchCenter2017}, first of which was that ``\emph{Public support for same-sex marriage has grown rapidly over the past decade}''.  The report looked at 34 separate nationally representative surveys conducted from 2004 to 2017 which each asked respondents whether they favor or oppose allowing gays and lesbians to marry (or don't know).  We replicate this study by using differential privacy to release the mean favor and oppose rates for each survey.  The true and DP survey means are shown in Figure~\ref{fig:pew} in blue and red respectively.  We used a conservative $\epsilon\!=\!0.01$ for each survey mean; while the mean and median sample sizes across surveys are slightly over 2000 respondents, the range in sample size is relatively large, giving different degrees of noise across released means.  Although each of the released means are noisy, the expected error is zero, and the trend remains.  As shown clearly in the figure, the trend from the best fit line across the DP releases is very close to the trend line from using the raw data, and an analyst exploring this data with \thesystem\ could have discovered the same finding with a high level of privacy protection.

\newpage
\bibliographystyle{plain}
\bibliography{TPDP}
\end{document}